\documentclass[9pt, conference]{IEEEtran}
\ifCLASSINFOpdf
\else
\fi
\usepackage{booktabs} 

\usepackage{amsfonts}

\usepackage[noend]{algpseudocode}
\usepackage{algorithm}

\usepackage{amsmath}
\usepackage{wrapfig}
\usepackage{floatflt}
\usepackage{diagbox}
\usepackage[abs]{overpic}
\usepackage{subcaption}

\usepackage{tabulary}
\usepackage{tabularx}
\usepackage[textsize=scriptsize]{todonotes}
\usepackage{caption}

\usepackage{multirow}

\usepackage{pgfplots}
\usepackage{tikz}
\usetikzlibrary{
	arrows.meta,
	colorbrewer,
	decorations.pathreplacing,
	automata, 
	positioning }

\usepackage{kbordermatrix}

\usepackage{balance}

\usepackage{fancyhdr}

\newcommand{\framework} {pSPICE}
\algnewcommand{\LeftComment}[1]{\Statex \(\triangleright\) #1}

\usepackage[most]{tcolorbox}
\fancypagestyle{firstpage}{%
	\rhead{}
	\lhead{}
	\cfoot{	
		\begin{tcolorbox}[colback=white,
			colframe=red,
			width=17cm,
			arc=3mm, auto outer arc,
			]
			{\color{red}
				\copyright ~IEEE, [2019]. This is the author's version of the work. It is posted here by permission of IEEE for your personal use. Not for redistribution. The definitive version  will be published in the Proceedings of IEEE BigData 2019.}
		\end{tcolorbox}
	}
}

\begin{document}
%
\title{\framework: Partial Match Shedding for Complex Event Processing}


\author{\IEEEauthorblockN{Ahmad Slo, Sukanya Bhowmik, Albert Flaig, Kurt Rothermel}
\IEEEauthorblockA{Institute for Parallel and Distributed Systems\\
University of Stuttgart, 
Stuttgart, Germany\\
firstName.lastName@ipvs.uni-stuttgart.de}
}


%


\IEEEoverridecommandlockouts
\IEEEpubid{\makebox[\columnwidth]{978-1-7281-0858-2/19/\$31.00~\copyright2019 IEEE \hfill} \hspace{\columnsep}\makebox[\columnwidth]{ }}

\maketitle
\IEEEpubidadjcol

\thispagestyle{firstpage}
\pagestyle{plain}

\begin{abstract}
Complex event processing (CEP) systems continuously process input event streams to detect patterns. Over time, the input event rate might fluctuate and overshoot the system's capabilities. One way to reduce the overload on the system is to use load shedding. 
In this paper, we propose a load shedding strategy for CEP systems which drops a portion of the CEP operator's internal state (a.k.a. partial matches) to maintain a given latency bound.
The crucial question here is how many and which partial matches to drop so that a given latency bound is maintained while minimizing the degradation in the quality of results.
In the stream processing domain, different load shedding strategies have been proposed that mainly depend on the importance of individual tuples. However, as CEP systems perform pattern detection, the importance  of events is also influenced by other events in the stream.
Our load shedding strategy uses Markov chain and Markov reward process to predict the utility/importance of partial matches to determine the ones to be dropped. In addition, we represent the utility in a way that minimizes the overhead of load shedding.
Furthermore, we provide algorithms to decide when to start dropping partial matches and how many partial matches to drop. 
By extensively evaluating our approach on three real-world datasets and several representative queries, we show that the adverse impact of our load shedding strategy on the quality of results is considerably less than the impact of state-of-the-art load shedding strategies. 

\end{abstract}

\vspace{0.4cm}


%
\IEEEpeerreviewmaketitle

\setlength{\textfloatsep}{-0pt}
\setlength{\intextsep}{0pt}

\section{Introduction}
\label{sec:introduction}
Complex event processing (CEP) is a powerful paradigm to detect patterns in continuous input event streams. The application area of CEP is very broad, e.g., transportation, stock market, network monitoring, game analytics, retail management,  etc. \cite{ Zacheilas:2015, spectre:2017, 1:Balkesen:2013:RRI:2488222.2488257,  3:Olston:2003:AFC:872757.872825, 1:debs2013, Wu:2006:HCE:SASE}. A CEP operator performs pattern matching by correlating the input events (also called primitive events) to detect important situations (called complex events) \cite{ Wu:2006:HCE:SASE, 1:Balkesen:2013:RRI:2488222.2488257, spectre:2017}.

In many applications, e.g., network monitoring, traffic monitoring, stock market \cite{3:Olston:2003:AFC:872757.872825, Zacheilas:2015, 1:Balkesen:2013:RRI:2488222.2488257}, the volume of input event streams is too high and it is not feasible to process the incoming events on a single machine. Moreover, the detection latency of complex events is significantly important, where the detected complex events might be useless if they are not detected within a certain latency bound \cite{Quoc:2017:SAC:3135974.3135989, CastroFernandez:2013:ISO:2463676.2465282}.
A well-known solution in CEP systems to process such  huge input event streams and maintain a defined latency bound is by using parallelization where the CEP operator graph is parallelized on multiple compute nodes. However,  the volume of input event streams is not stable and fluctuates over time \cite{8622265:concept-drift, 3:Rivetti:2016:LSS:2933267.2933311}. Therefore, it is not trivial to know the number of necessary compute nodes in advance. Hence, either the number of compute nodes should be over-provisioned, which  introduces additional cost, or the number of compute nodes can be adapted elastically as proposed by many researchers \cite{1:Balkesen:2013:RRI:2488222.2488257,  1:Zeitler413076, CastroFernandez:2013:ISO:2463676.2465282, Zacheilas:2015, 1:s4:5693297}.
However, adapting the  parallelization degree in case of short input spikes introduces a high performance overhead \cite{8622265:concept-drift}. Moreover, resources might be limited  for several reasons: 1)  limited monetary budget, 2) limited compute resources if operators run in private clouds due to security or response time reasons.

Therefore, to handle system overload in case of limited resources or in case of short input event spikes, load shedding might have to be used. 
Load shedding has been extensively studied in the stream processing domain \cite{3:Tatbul:2003:LSD:1315451.1315479, 3:Tatbul:2006:WLS:1182635.1164196, 3:Rivetti:2016:LSS:2933267.2933311, 3:Olston:2003:AFC:872757.872825}.
The queries in this domain (e.g., aggregation, min, max, etc.) depend on individual tuples and hence researchers propose approaches to assign utilities to the  tuples individually without taking into consideration the dependency between tuples. However, CEP systems perform pattern correlation operations between different events.  Hence, we must take into consideration the dependency between events in patterns. 
In \cite{3:He2014OnLS}, the authors proposed a load shedding strategy for CEP systems where they assign utilities to events depending on the dependency between events in the patterns and accordingly shed events. However, they do not consider the order of events in patterns which is important in CEP as in sequence  and negation operators \cite{4812454, Cadonna:2011:SES:1951365.1951372}. In \cite{eSPICE}, the authors proposed a load shedding strategy, called eSPICE, for CEP systems. eSPICE drops events from the operator's input event stream where it takes into consideration the dependency between events and their order in patterns. 

Both the aforementioned load shedding strategies in CEP \cite{3:He2014OnLS, eSPICE} use a black-box approach where primitive events are dropped from the input event queue of a CEP operator. However, load shedding may be performed in CEP using a white-box approach as well where a portion of the operator's internal state is dropped.
Of course, dropping might adversely impact the quality of results (denoted by QoR), i.e., detected complex events, where important situations could be missed.
Moreover, using a black-box approach to drop primitive events may introduce falsely detected complex events, e.g., when using negation operator.  Therefore, it is crucial to shed load in a way that has low impact on QoR. 
In this paper, we propose a white-box load shedding approach and compare it with state-of-the-art black-box approaches \cite{3:He2014OnLS, 3:Tatbul:2003:LSD:1315451.1315479} to show the advantages of a white-box approach.

More specifically, we propose an efficient load shedding strategy for CEP systems, called \framework, that considers event dependency and order in patterns. \framework~drops a portion of the internal state of a CEP operator. The internal state contains information about  partial matches, where a \textit{partial match} is a detected part of a pattern which could become a \textit{complex event} if the full pattern is matched. As a short hand, we call information about a partial match in the internal state of a CEP operator as a partial match (PM) hereafter.
The event processing latency increases proportionally with number of PMs in an operator \cite{Ray:2016:SPS:2882903.2882947, 1:Balkesen:2013:RRI:2488222.2488257}. Therefore, dropping PMs from an operator reduces the event processing latency and increases the operator throughput. Hence, it enables the operator to maintain a defined latency bound in case of input event overload.
\framework~drops PMs that have low adverse impact on  QoR.

There are three main challenges to drop partial matches in CEP: 
1) determining when and how many PMs to drop for an incoming input event rate,
2) determining which PMs to drop, and
3) performing the load shedding in a light-weight  manner so as not to burden an already overloaded operator.
To drop PMs, we associate each PM with a utility value that indicates the importance of the PM where a higher utility value means a higher importance. We derive the utility of a PM from its probability to complete and become a complex event (called partial match completion probability) and from its estimated remaining processing time. 

Our contributions are as follows: 
\begin{itemize}
	\item 
	We propose a new load shedding strategy, called \framework, that uses Markov chain and Markov reward process to predict the utility of PMs in windows. 
	The utility of a PM depends on the completion probability of the PM and on its remaining processing time.
	
	\item We develop an approach that enables us to perform the load shedding in an efficient and light-weight manner.
	\item
	 We provide an algorithm that decides when and estimates how many PMs to drop from an operator to maintain the given latency bound.

	 \item
	 We provide extensive evaluations on three real-world datasets and several representative queries to show that \framework~reduces the adverse impact of load shedding on QoR considerably more than state-of-the-art solutions.
\end{itemize}

\section{Preliminaries and Problem Statement}
\label{sec:background}

\subsection{Complex Event Processing}
CEP systems process input event streams to detect patterns. A CEP system may consist of one or more operators represented by a directed acyclic graph (DAG), called operator graph. Each operator processes input event streams originating from several sources to detect a set of patterns. Sources could be sensors, upstream operators, other applications, etc.
An event (also called primitive event) in the input event stream consists of attribute-value pairs. The attribute-value pairs contain the event data, e.g., stock quote in a stock application, player position in a soccer application, or bus location in a transportation application. The attribute-value pairs might also contain sequence number and/or timestamp. Events in the input event streams have global order, for example, by using the sequence number or the timestamp and a tie-breaker.

In this work, we focus on a CEP system consisting of a single CEP operator where the operator might detect one or more patterns (i.e., multi-query). 
To detect important situations (complex events), an operator processes primitive events in the input event stream and matches a set of patterns $\mathbb{Q}= \{q_1, q_2, ...,q_n\}$, where n represents number of patterns. Patterns might have different importances and hence each pattern has a corresponding weight (given by the domain expert) that indicates its importance. The pattern weight $\mathbb{W_Q}$ is defined as: $\mathbb{W_Q}= \{w_{q_1}, w_{q_2}, ...,w_{q_n}\}$, where $w_{q_x}$ is the weight of pattern $q_x$. A pattern in CEP is defined using an event specification language like Tesla \cite{1:cu2010tesla} or SASE \cite{Wu:2006:HCE:SASE}. These languages contain several types of CEP operators: sequence, conjunction, negation, etc.

The input event stream is continuous and infinite, however in CEP, the input event stream is partitioned using predicates into independent chunks of events, called windows. Windows capture the temporal relationship between the primitive events in the input event stream.  The predicates to open and close windows may depend on time
(called time-based window), on the number of events (called count-based window), on logical predicates (called pattern-based window), or on a combination of them \cite{1:Balkesen:2013:RRI:2488222.2488257, 3:Tatbul:2006:WLS:1182635.1164196}.

To detect patterns, a CEP operator performs pattern matching using a \textit{process} function that processes the incoming windows of primitive events and searches for patterns within these windows. Windows may overlap, however, they are processed independently by the \textit{process} function.
In a window $w$, a matched part of a pattern is called a \textit{partial match} (PM). We define a partial match $pm$ of a pattern $q_x \in \mathbb{Q}$ as follows: $pm \subset q_x$. In the window $w$, the partial match $pm \subset q_x$ becomes a complex event if the pattern $q_x$ is completely matched. PMs represent a part of an operator's internal state.

In CEP systems, the pattern matching operation can be represented as a finite state machine \cite{Ray:2016:SPS:2882903.2882947, spectre:2017, 1:Balkesen:2013:RRI:2488222.2488257}, where a partial match represents an instance of this state machine. For a pattern $q_x \in \mathbb{Q}$, we define a set of states $\mathbb{S}_{q_x} = \{s_1, s_2, ..., s_m\}$ as a set of all possible states that the pattern $q_x$ can have, including the initial state ($s_1=\phi$). 
Let us look at an example of a pattern to understand its properties.  In a traffic monitoring system \cite{Zacheilas:2015}, if more than one bus has delay at the same bus stop, it might indicate an abnormal traffic, e.g., an accident. To detect the abnormal behavior (i.e., a pattern), a traffic analyst formulates the following query using the Tesla  \cite{1:cu2010tesla} event specification language:
\begin{equation*}[q_e]
	\resizebox{0.90\linewidth}{!}{$
	\begin{aligned}
		\ & \mathtt{define\ Abnormal(Bus1, Bus2, Bus3)} \\
		\ & \mathtt{from\  BusEvent\ (delay > \$x) ~as~ e_A \ and } \\
		\ & \mathtt{BusEvent() ~as~ e_B ~and~ (delay> \$x\ and \ e_B.stop = e_A.stop)}\\
		\ &\ \mathtt{ \qquad \qquad within\ 5min\ from\ e_A \ and}\\
		\ & \mathtt{BusEvent() ~as~ e_C ~and~ (delay> \$x\ and \ e_C.stop = e_A.stop)}\\
		\ &\ \mathtt{ \qquad \qquad within\ 5min\ from\ e_A}\\
		\ & \mathtt{where\ Bus1= e_A.Id,\ Bus2=e_B.Id,\ Bus3=e_C.Id}
	\end{aligned}
$}
\end{equation*}

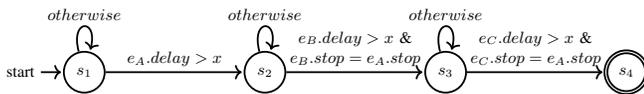
\begin{figure}
	\centering
	\resizebox{0.99\linewidth}{!}{

\tikzset{label style/.style={font=\fontsize{24}{26}\selectfont\bf, color=black}
	 }

\begin{tikzpicture}[shorten >=1pt,node distance=3.5cm,on grid,auto,  line width=1pt] 
\node[state,initial] (s_1)   {$s_1$}; 
\node[state] (s_2) [right=of s_1] {$s_2$}; 
\node[state] (s_3) [right=of s_2] {$s_3$}; 
\node[state, accepting] (s_4) [right=of s_3] {$s_4$}; 

\path[->] 
(s_1) edge  node[align=center] {$e_A.delay > x$} (s_2)
	 edge [loop above] node {$otherwise$} ()
(s_2) edge  node[align=center] {$e_B.delay > x$ \& \\  $e_B.stop = e_A.stop$} (s_3)
	edge [loop above] node {$otherwise$} ()
(s_3) edge  node[align=center] {$e_C.delay > x$ \& \\ $e_C.stop = e_A.stop$} (s_4)
	edge [loop above] node {$otherwise$} ()	
;
\end{tikzpicture}}
	\caption{State machine for $q_e$}
	\label{fig:bus-state-example}
\end{figure}

$q_e$ detects an abnormal traffic, i.e., a complex event, if a bus event $e_A$ is delayed on a certain stop and the following bus events $e_B$ and $e_C$ within 5 minutes (window size) from bus event $e_A$ also get delayed on the same stop. Figure \ref{fig:bus-state-example} shows the corresponding state machine of $q_e$.
With each incoming bus event $e_A$, a new window $w$ is opened. In addition, an instance of the state machine of $q_e$ is created with the initial state $s_1$.  If the bus event $e_A$ indicates that the bus is delayed, a new partial match $pm$ is opened and the state machine progresses to the next state. This is shown in Figure \ref{fig:bus-state-example}, where the state machine transitions from the initial state ($s_1$) to the state  $s_2$. In the window $w$, with each subsequent bus event $e_B$, $pm$ progresses towards completion if $e_B$ indicates that the subsequent bus is also delayed at the same stop as $e_A$, i.e., the state machine transitions from the state $s_2$ to the state $s_3$. When receiving the bus event $e_C$, $pm$ completes, i.e., becomes a complex event, if the bus event $e_C$ indicates that the subsequent bus is also delayed at the same stop as $e_A$, i.e., the state machine transitions to the state $s_4$ (the final state).

In this example, the number of states in the state machine is 4, including the initial state ($s_1=\phi$), i.e., $|\mathbb{S}_{q_e}| = 4$. Please note that, in $q_e$, the bus event $e_A$ has less processing latency than the bus events $e_B$ and $e_C$ since the bus events $e_B$ and $e_C$ must be checked with more conditions than the bus event $e_A$. Hence, events in a pattern might have different processing latencies.

In this paper, the only assumption that we have is that operators reveal information about the progress of PMs when processing primitive events within windows.

\subsection{Problem Statement}
As discussed earlier, in this paper, we drop PMs from the operator's internal  state to maintain a given latency bound ($LB$) in the face of system overload.  However, dropping PMs  might degrade QoR, where quality is measured by the number of missed complex events, i.e., false negatives. Since we do not drop primitive events but PMs, our load shedding strategy does not result in producing false positives, i.e., detect a complex event which should not be detected.  
To minimize its impact on QoR, the load shedder must drop only those PMs that have low utility/importance.
The utility of PMs is measured by their influence on the number of \textit{missed complex events}, i.e., false negatives.

As mentioned above, an operator matches a set of patterns $\mathbb{Q} = \{q_1, q_2, ...,q_n\}$ and, therefore, QoR is measured by the sum of false negatives (denoted by $FN_\mathbb{Q}$) of all patterns in the operator.
We define the number of false negatives for each pattern $q_x$ as $FN_{q_x}$. Since each pattern $q_x$ has a weight $w_{q_x}$, which indicates its importance,  we must also  consider the pattern's weight $w_{q_x}$ when calculating $FN_\mathbb{Q}$. 

QoR is formally defined as follows:
\begin{equation*}
\begin{aligned}
\ & \ FN_\mathbb{Q} = \sum_{i=1}^{n} w_{q_i} . FN_{q_i}
\end{aligned}
\end{equation*}
The objective is to minimize $FN_\mathbb{Q}$, such that the given latency bound $LB$ is met.

\section{\framework}
\label{sec:ls}
In this section, we first present  the architecture of \framework, our load shedding strategy. Then, we introduce the notion of utility of PMs followed by a description of our approach to determine these utilities using Markov chain and Markov reward process \cite{howard1971dynamic}. After that, we discuss how to detect overload and compute the amount of overflowing PMs that must be dropped by the load shedder. Finally, we present the load shedding algorithm which efficiently drops PMs with the lowest utility values.

\subsection{Load Shedding Architecture}
The architecture of \framework~is depicted in Figure \ref{fig:ls_architecture}. The figure shows an operator which is modified by adding the following components to enable load shedding: overload detector, load shedder (LS) and model builder. 

The incoming windows of events forwarded by an upstream operator (e.g., window operator) are queued in the input queue of the operator. To prevent violating the defined latency bound ($LB$), the overload detector  checks the estimated latency for each input event. In the scenario where $LB$ might be violated, the overload detector calls the LS to drop a certain number of PMs, denoted by $\rho$.

The model builder receives observations from the operator about the progress of PMs. After receiving a certain number of observations, the model builder builds the model, where it predicts the utility of PMs using Markov chain and Markov reward process. The model builder might be heavy-weight. However, it is not a time-critical task and it does not need to run frequently.

LS drops $\rho$ PMs every time it is called by the overload detector, where $\rho$ is determined by the overload detector. The LS depends on utility values predicted by the model builder to select those PMs for dropping. Both the LS and overload detector are time-critical tasks where they directly affect the CEP system performance and hence they must be light-weight and efficient. As we will see later, both of these components have very low overhead in \framework.

\begin{figure}[t]
	\centering
	\resizebox{0.99\linewidth}{!}{

\newcommand{\gear}[6]{%
  (0:#2)
  \foreach \i [evaluate=\i as \n using {\i-1)*360/#1 }] in {1,...,#1}{%
   arc (\n:\n+#4:#2) {[rounded corners=1.5pt] -- (\n+#4+#5:#3)
    arc (\n+#4+#5:\n+360/#1-#5:#3)} --  (\n+360/#1:#2)
  }%
  (0,0) circle[radius=#6] 
}

\tikzset{label style/.style={font=\fontsize{24}{26}\selectfont\bf, color=black},
	rectangle style/.style ={color= blue, line width=4pt }, 
	arrow style/.style={color=black, line width=2pt, font=\fontsize{24}{26}\selectfont\bf, -{>[scale=2.5, length=7, width=4]}}, 
	instance style/.style={color=blue, line width=3pt} }

\begin{tikzpicture}[]

\draw  [arrow style, label style ] (2.5, 3) -- (5.5,3) node[pos=0.4, below, align= left] {windows};

%

\newcommand\XStart{5.5}
\newcommand\YStart{2.5}
\newcommand\SquareSize{1}
\foreach \n in {0,...,4}
	\draw[rectangle style, rounded corners] (\n + \XStart, \YStart) rectangle (\n + \XStart + \SquareSize, \YStart+ \SquareSize);
	
\node [label style ] at (8, 4) {input queue};

\draw  [arrow style, label style ] (10.5, 3) -- (11.5, 3);

\draw[rectangle style] (11.5,2) rectangle (15.1,4)  node [pos=.5,  label style, align=left]{overload\\detector};
\draw  [arrow style, label style ] (12.75, 2) -- (11.75, -2.25) node[pos=0.5,above, sloped] {drop($\rho$)};

\draw  [arrow style, label style ] (15.1, 3) -- (17, 3);	
	
\draw [instance style](17.75, 3 ) ellipse [x radius=2.2cm,y radius=2cm]; 
\node[label  style, above] at (17.75,5) {operator};
\begin{scope}[xshift=17.75cm, yshift=3cm]
\fill[even odd rule]  \gear{8}{0.5}{0.75}{10}{2}{0.25};
\end{scope}
\node[label  style, above] at (17.75, 3.6) {process};
\node[label style, rectangle, draw, blue, line width=2pt, minimum width=0.5cm,minimum height=0.5cm, text= black] (PMs) at (17.75, 1.75) {PMs}; 

\draw  [arrow style, label style ] (18.5, 3) -- (22.5, 3) node[pos=0.8, above=2pt, align= left] {complex\\ events};

\draw[rectangle style ] (16.5,-5) rectangle (21.8, -2)  node [pos=.5, above= 12pt, label style, align=left]{model builder};
\draw[rectangle style ] (17.5,-4.5) rectangle (20.8, -3.2)  node [pos=.5,  label style, align=left]{model};
\draw[arrow style, label style ] (PMs) --  (19, -2) node[ pos= 0.5, above=0.1, sloped]{observ.} ;

\draw [rectangle style, red] (11, -2) -- (12.5, -2.75) -- (12.5, -3.25) -- (11, -4) --cycle;
\node[label  style] at (11.75, -3) {LS};
\draw  [arrow style, label style ] (17.5, -3.80) -- (12, -3.5) node[pos= 0.4, below,  sloped] {$\mathbf{U_{pm}}$};
\draw  [arrow style, label style ] (12.5, -3) -- (PMs) node[pos=0.5,above,  sloped] {getPMs()};
\draw  [arrow style, label style ] (PMs) --(12.5, -3)  node[pos=0.5,below,  sloped] {removePM(Id)};

\end{tikzpicture}}
	\caption{\framework~Architecture.}
	\label{fig:ls_architecture}
\end{figure}

\subsection{Utility of Partial Matches}
\framework~drops partial  matches with the lowest utility/importance. 
The question is-- what defines  the utility of a PM? The utility/importance of a PM is defined by its impact on QoR, i.e., number of false negatives. A PM that has a low adverse impact on QoR has a low utility value, while a PM that has a high adverse impact on QoR has a high utility value.
Hence, to minimize the dropping impact on QoR, we must find a way to assign low utility values to those PMs that are less important than other PMs.
We assign utilities to PMs depending on three factors: 1) the probability of a PM to complete and become a complex event (i.e.,  the completion probability), 2) the estimated processing time that a PM still needs, and 3) the weight of the pattern.

The completion probability of a PM represents the probability of the PM to become a complex event.
The existence of a complex event depends on whether its underlying PM will complete or not. If a PM completes, a complex event is detected. On the other hand, if a PM does not complete, a complex event is not detected. Hence, the completion probability of a PM is an important indicator of the utility/importance of the PM as dropping PMs that in anyway will not complete implies no degradation in QoR.  
$P_{pm}$ represents the completion probability of the PM $pm$. 
Higher is the completion probability of the PM  ($P_{pm}$), higher should be its utility. This means that the utility of a PM is proportional to its completion probability.   

The utility of a partial  match $pm$ is also  influenced by its remaining processing time (denoted by $\tau_{pm}$). A PM that still has a high remaining processing time (we will use only processing time hereafter) should have lower utility than  a PM that has a lower processing time. The reason for this is that a PM with low processing time consumes less processing time from the operator, i.e., giving the operator more time to process other PMs. Hence, it decreases the need to drop PMs from the operator's internal state, which in turn decreases the number of false negatives.
This means that the utility of a PM is inversely proportional to its processing time ($\tau_{pm}$).

For example, let us assume that an operator has two partial matches $pm_1$ and $pm_2$ in two windows $w_1$ and $w_2$, respectively. Suppose that $P_{pm_1}= P_{pm_2}$  but $\tau_{pm_1} < \tau_{pm_2}$. In this case, the importance of $pm_1$ should be higher than the importance of $pm_2$ since $pm_1$ has the same completion probability as $pm_2$ but it imposes lower processing time on the operator. 
In another case where $P_{pm_1} < P_{pm_2}$  but $\tau_{pm_1} < \tau_{pm_2}$, we need to assign a higher utility to the PM that results in lesser degradation in QoR. Therefore, we use the proportion of the completion probability $P_{pm}$ to the processing time $\tau_{pm}$, i.e., $\frac{P_{pm}}{\tau_{pm}}$, as a utility value for the partial match.

Finally, as we mentioned above, in an operator with multiple patterns (i.e., multi-query operator), each pattern might have different weight $w_{q_x}$, i.e., different importance. Therefore, when assigning utilities to PMs, we must also take the patterns' weights into  consideration. To consider the pattern's weights, we increase the utility value of a PM $pm  \subset q_x$ proportionally to its pattern's weight $w_{q_x}$.

To incorporate the completion probability of a PM $P_{pm}$, its processing time $\tau_{pm}$, and its pattern's weight  $w_{q_x}$ in deriving the utility of the PM (denoted by $U_{pm}$), we represent the utility of a PM as follows: 

\begin{equation}
	U_{pm}= w_{q_x} . \dfrac{P_{pm}}{\tau_{pm}}
	\label{eq:ut}
\end{equation}


\subsection{Utility Prediction}
\label{sec:utility}
Since the utility of a PM  depends on its completion probability and processing time, in this section, we explain the manner in which we predict them using Markov chain and Markov reward process.

\subsubsection{Completion probability Prediction}
In a certain  position in a window $w$, the completion probability $P_{pm}$ of  a PM $pm \subset q_x$, i.e., the probability of $pm$ to complete the pattern $q_x$ and to become a complex event, depends on two factors. 1) on \textbf{the current state} of the PM $pm$ (denoted by $S_{pm}$), where $S_{pm}  \in \mathbb{S}_{q_x}$, and 2) on \textbf{ the number of remaining events} in the window $w$ (denoted by $R_w$). Therefore, we write $P_{pm}$ as a function of $S_{pm}$ and $R_w$ as follows: 

\begin{equation}
	P_{pm}= f(S_{pm}, R_w)
\end{equation}

$ S_{pm}= s_1$ means that $pm$ is in the initial state while $ S_{pm}=  s_m$, where $m = |\mathbb{S}_{q_x}|$, means that $pm$ has completed and become a complex event.
$R_w \in [1, ws]$, where $ws$ represents the expected window size.
In case a partial match $pm$  has a state $ S_{pm}$ which is close to the final state and $R_w$ is high,  the probability for $pm$ to complete $q_x$ and become a complex event might be high. This is because $pm$ needs only fewer state transitions to reach the final state and  the window $w$ still has a high number of events that can be used to match the pattern $q_x$ and to complete $pm$. 
On the other hand,  the completion probability might be low for a partial match $pm$  that has a state $ S_{pm}$ which is close to the initial state and $R_w$ is low. This is because $pm$  still needs many state transitions to reach the final state and  the window $w$ only has a small number of events that can be used to match the pattern $q_x$ and to complete $pm$.

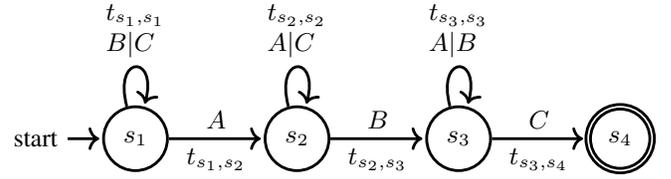
\begin{figure}
	\centering
	\resizebox{0.99\linewidth}{!}{

\begin{tikzpicture}[shorten >=1pt,node distance=2cm,on grid,auto, line width=1pt] 
\node[state,initial] (s_1)   {$s_1$}; 
\node[state] (s_2) [right=of s_1] {$s_2$}; 
\node[state] (s_3) [right=of s_2] {$s_3$}; 
\node[state, accepting] (s_4) [right=of s_3] {$s_4$}; 


\path[->] 
(s_1) edge  node[] {$A$}  node[below] {$t_{s_1,s_2}$}(s_2)
edge [loop above] node[align=left] {$t_{s_1,s_1}$ \\ $B|C$} ()
(s_2) edge  node {$B$} node[below] {$t_{s_2,s_3}$} (s_3)
edge [loop above] node[align=left] {$t_{s_2,s_2}$ \\$A|C$} ()
(s_3) edge  node {$C$} node[below] {$t_{s_3,s_4}$}(s_4)
edge [loop above] node[align=left] {$t_{s_3,s_3}$ \\$A|B$} ()	
;

\end{tikzpicture}}
		\caption{State machine example}
		\label{fig:state-example}
\end{figure}

Since a pattern in CEP systems can be represented as a state machine, as we mentioned above, in this work, we model the pattern matching as a Markov chain to predict the completion probability of a partial match $P_{pm}$ of a pattern. To clarify this, let us introduce the following simple example. Let us assume that an operator matches a pattern $q_x= seq (A;B;C)$.  
This pattern can be represented as a state machine as depicted in Figure \ref{fig:state-example}, where it has four states, including the initial state, i.e., $\mathbb{S}_{q_x}=\{s_1, s_2, s_3, s_4\}$. 
The state machine transitions from one state to other states depending on the input symbols (events), while the Markov chain probabilistically transitions from one state to other states using a \emph{transition matrix}. 
In the above example, if we assume that the input event stream has only three event types (A, B, and C) and these events are coming randomly with a uniform distribution, then the probability to transition from any state to next state is 1/3. While the probability to stay in the same state is 2/3. 

Therefore, the transition matrix can be used to predict the probability of the state machine to transition from any state to other states and hence to predict the  probability of the state machine to transition from a certain state $s_i$ to the final state $s_m$ after processing  $R_w$ input symbols.
Since a PM is represented as an instance of the state machine, the completion  probability of a PM (i.e., the probability of the state machine to reach the final state) in a certain state $S_{pm}$ given that $R_w$ events are left in the window $w$, i.e., $P_{pm}= f(S_{pm}, R_w)$, can be computed using the transition matrix.
Since the input event stream might follow any distribution, not only uniform distribution, we should learn the transition matrix by gathering statistics about the state transitions of PMs as we describe next.

\textbf{Statistic gathering \& Transition matrix: }
For each pattern $q_x \in \mathbb{Q}$, the model builder builds a transition matrix $T_{q_x}$ from the statistics gathered  during run-time by monitoring the internal state of the operator.  The statistics contain information about the progress of PMs of pattern $q_x$ when processing the input events within windows.
For each partial match $pm \subset q_x$, the operator reports, when processing an input event $e$ within a window, whether $pm$ progressed or not, i.e., the state of $pm$ changed or not by processing the event $e$. The operator forms an $Observation<q_x, s, s'>$, where $s$ represents the current state of $pm$ and $s'$ the new state of $pm$ after processing one event in the window.

After gathering statistics from $\eta$ observations for pattern $q_x$, the model builder transfers these statistics to the transition matrix $T_{q_x}$. $T_{q_x}$ describes the \textit{transition probability} between the states of  Markov chain when processing \emph{one event} in a window.

\textbf{Completion probability:}
As mentioned above, the transition matrix gives the probability to transition from one state to another state and can be used to predict the partial match completion probability $P_{pm}$.  Figure \ref{fig:transition-matrix} shows the transition matrix for the state machine depicted in Figure \ref{fig:state-example}.
Since we are only interested to know whether a partial match will complete or not, we need only to focus on the last column in the transition matrix, surrounded with a red box in Figure \ref{fig:transition-matrix}. This column gives the probability to move from any state $S_{pm}$  to the final state, i.e., the probability to complete the partial match.

The transition matrix contains the transition probability, given that there is only one event left in a window. Therefore, to get the transition probability given that $R_w$ events are still left in a window $w$, we must raise the transition matrix $T_{q_x}$ to the power $R_w$. This way, the completion probability of a partial match $pm$ in a state $S_{pm}$, given that $R_w$ events are left in a window $w$, is computed as follows: 

\begin{equation}
	P_{pm}= f(S_{pm}, R_w)= T_{q_x}^{R_w}(i, m )
\end{equation}
where $S_{pm}= s_i \in \mathbb{S}_{q_x}$  and $m= |\mathbb{S}_{q_x}|$, $m=4$ in the above figure. 
For example, in Figure \ref{fig:transition-matrix}, the completion probability of a partial match $pm \subset q_x$ in the state $s_2$ given that only one event ($R_w=1$) is left in a window $w$ is computed as follows: $P_{pm} = f(s_2, 1)= T_{q_x}^{1}(2, 4)= p_{24}$.
To get the completion probability of a partial match given any number of events are left in a window, we  need to compute the transition matrix $T_{q_x}^{R_w}$ for all possible values of $R_w \in [1, ws]$.
However, the window size $ws$ might be too large which might impose a high memory cost during the calculation of the transition matrices. Therefore, we calculate the transition matrix only for every $bs$ (i.e., bin size)  events, i.e., $T_{q_x}^{bs}$, $T_{q_x}^{2.bs}$, ..., $T_{q_x}^{ws}$.  To get the completion probability of a PM in case $R_w \in [(j- 1).bs,~j.bs]$, where $j=1, 2, ..., \frac{ws}{bs}$, we use linear interpolation. For ease of presentation, we assume that $bs=1$, if not otherwise stated.

\begin{figure}[t]
	\centering
\begin{tabular}{c@{}c@{}}
	& \mbox{\hspace{4mm}next state} \\ 
	\parbox[c][15mm][t]{3mm}{\rotatebox{90}{current state}} &
	$ \kbordermatrix{
		& s_1   & s_2   & s_3  & s_4  \cr
		s_1 & p_{11} & p_{12} & p_{13} & \smash{\color{red}\setlength{\fboxrule}{1.5pt}\fbox{\color{black}\rule[-33pt]{-2pt}{0pt} $p_{14}$ }} \cr
		s_2 & p_{21} & p_{22} & p_{23} & p_{24} \cr
		s_3 & p_{31} & p_{32} & p_{33} & p_{34} \cr
		s_4 & p_{41} & p_{42} & p_{43} & p_{44} \cr
	} $ \\ 
\end{tabular}
\caption{Transition matrix $T_{q_x}$ for the state machine in Figure \ref{fig:state-example}.}
\label{fig:transition-matrix}
\end{figure}

\subsubsection{Processing Time Prediction}
After predicting the completion probability of PMs, now, we describe how to predict the processing time of PMs using Markov reward process, where we model the processing time of a PM as the reward value.
Given a partial  match $pm \subset q_x$ in  a state $S_{pm}$,  we define the time that is needed to match an event in a window $w$ with $pm$ as $t_{s,s'}$, where $s=S_{pm}$ and $s' \in \mathbb{S}_{q_x} $. Hence, $t_{s,s'}$ represents the processing time that is needed for the state machine of $q_x$ to transition from state $s$ to state $s'$. For example, in Figure  \ref{fig:state-example}, the processing time to transition from $s_1$ to $s_2$ is represented by the value $t_{s_1,s_2}$. 
We consider $t_{s,s'}$ as a reward value to move from state $s$ to state $s'$ in the state machine of $q_x$.

Therefore, to calculate the processing time of a PM, we clearly need something more than using Markov chain which is used to compute the completion probability of a PM. As a result, we upgrade our Markov chain to Markov reward process, where we additionally define the reward function $R_{q_x}(s,s')$ as the expected processing time needed to transition from state $s$ to state $s'$.
Solving Markov reward process gives us the expected reward for each state in the state machine, given that there are still $R_w$ events left in a window $w$. 
Since we represent the processing time $t_{s,s'}$ as reward, the reward of a state represents the estimated processing time of a PM $\tau_{pm}$,  given that there are still $R_w$ events lefts in a window $w$.

We incorporate the processing time $t_{s,s'}$ in statistics gathering and extend the above observation as follows: $Observation<q_x, s, s', t_{s,s'}>$. 
After gathering statistics from $\eta$ observations for pattern $q_x$, the model builder constructs the reward function (i.e., $R_{q_x}(s, s')$) which is calculated as the average value for all observed values of the processing time $t_{s,s'}$. After that, the model builder predicts the processing time of PMs by solving Markov reward process as we explain next.

\textbf{Processing time:}
To predict the processing time of a partial match $pm \subset q_x$ in a state $S_{pm}$, the model builder must solve Markov reward process.
A well-known algorithm called \textit{value iteration} \cite{howard1971dynamic}  can be used to solve Markov reward process.  The algorithm iteratively calculates the expected reward (processing time $\tau_{pm}$ in this case) at every state in  the state machine using the transition matrix $T_{q_x}$ and the reward function $R_{q_x}$. Then, it reuses the calculated reward values in the future iterations. Here, an iteration $j$ represents the number of remaining events (i.e., $R_w$) in a window $w$, i.e., $j = R_w$.  The value iteration algorithm uses the bellman equation \cite{Bellman:1957} to predict the remaining processing time $\tau_{pm}$ of a partial match $pm \subset q_x$ at state $S_{pm}$ given that there are still $R_w$ events left in the window $w$. 

Similar to the completion probability, we run the value iteration algorithm to get the processing time  $\tau_{pm}$ of a partial match $pm$ for all expected remaining $R_w$ number  of events in a window $w$. To avoid the memory overhead in case of too large window size $ws$, again, we keep the value iteration results only for every $bs$ events. For the intermediate values, we  use linear interpolation.

\subsubsection{Utility calculation}
After describing how to predict the completion probability and the processing time of a PM, now, we can derive the utility of PMs for each pattern $q_x \in \mathbb{Q}$ using Equation (\ref{eq:ut}). 

Since the completion probabilities and  processing times of PMs have different units and scales, using Equation (\ref{eq:ut}) directly on these values, may result in unexpected behavior, where a high processing time may overcome the completion probability and eliminate its importance in calculating the utility of PMs. Therefore, before using Equation (\ref{eq:ut}), we bring the completion probabilities and processing times to the same scale
and then apply Equation (\ref{eq:ut}) to get utilities of PMs. 

To efficiently retrieve the utilities by the LS, we store the utility of PMs at any given state and for any number of remaining events in a window in a table called $UT_{q_x}$, where each pattern $q_x$ has its corresponding utility table. $UT_{q_x}$ has $(\frac{ws}{bs}~ X ~m)$ dimensions, where $m= |\mathbb{S}_{q_x}|$ and each cell $UT_{q_x}(i,j)$ represents the utility of a PM in state $s_i$ given that there are still $j$ events left in the window, assuming $bs=1$. So the utility of a PM $pm \subset q_x$ is calculated as follows: $U_{pm}= f(S_{pm}, R_w)= UT_{q_x}( i,j)$, where $s_i= S_{pm}$ and $j= R_w$.  
Getting the utility of a PM from $UT$ has only O(1) time complexity which is a great factor in minimizing the overhead of the LS.

\subsection{Model Retraining}
\label{sec:retrain}
The event distribution in the input event stream and/or the content of input events may change over time and hence our model might become inaccurate and adversely impact QoR. To avoid this, we must retrain the model to capture those changes.
The question is-- how do we know that those changes happened and the model must be retrained? We depend on the transition matrix to answer this question. 

The transition matrix, as we know, contains the probabilities to transition from any state to other states in the state machine, where the transition matrix is constructed depending on the distribution of input event stream and on the content of events. So, if there is a change in the distribution of input event stream and/or on the content of events, the probability values in the transition matrix will change.
Therefore, the transition matrix can be used as an indicator to those changes and to trigger model retraining. Hence, we propose to  periodically build a new transition matrix from the gathered statistics from the operator and compare the new transition matrix with the transition matrix that is used in the model by using an error measurement, e.g, mean squared error. If the deviation between the two matrices is higher than a threshold, the model builder must rebuild the model. Please note that building a new transition matrix is light-weight since we just need to transfer the gathered statistics about the state transitions to probability values. Moreover, we don't need to calculate new transition matrices for all expected remaining number of events in a window to check for the need to retrain the model.

\subsection{Detecting and Determining Overload} 
The goal of \framework~is to avoid violating a defined latency bound ($LB$). A high queuing latency of the incoming input events in the operator input queue indicates an overload on the operator and hence some partial matches must be dropped from the operator's internal state to avoid violating $LB$. 
Algorithm \ref{alg:overloadDetector} formally describes  the functionality of the overload detector .

\textbf{Detecting overload: } 
The overload detector continuously gets the primitive events from the event input queue of the operator, where for each event, it checks whether $LB$ might be violated. In the scenario where $LB$ might be violated, the overload detector calls the load shedder to drop a certain number of PMs to reduce the overhead on the operator and maintain $LB$.  
The violation of $LB$ depends on the estimated event latency (denoted by $l_e$) and load shedding latency (denoted by $l_s$), where $LB$ would be violated if the following inequality holds: 

\begin{equation}
	l_e + l_s > LB
	\label{eq:drop-condition}
\end{equation} 

The estimated event latency $l_e$  represents the time between the insertion of  the event $e$ in the operator's input queue and the time when the event $e$ is processed by the operator in all currently opened windows, since an event may belong to several windows in case windows overlap. 
The load shedding latency $l_s$ represents the time needed by the LS to drop the needed amount of partial matches. 

The estimated event latency $l_e$ of an event $e$ is the sum of the event queuing latency (denoted by $l_q$) and the estimated event processing latency (denoted by $l_p$): $l_e= l_q + l_p$. The event queuing latency $l_q$ is the time between the insertion of  the event $e$ in the operator's input queue and the time when the operator gets the event $e$ from its input queue to process it (cf. Algorithm \ref{alg:overloadDetector}, line 2). While, the estimated  event processing latency $l_p$, represents the time an event $e$ needs to be processed by the operator in all currently opened  windows. $l_p$ depends on current number of partial matches (denoted by $n_{pm}$) in the operator since the event $e$ needs to be matched with all current partial matches in the operator. Higher is the value of $n_{pm}$, higher is $l_p$. Therefore, we represent $l_p$ as a function, called event processing latency function, of the current number of partial matches $n_{pm}$ in the operator: $l_p= f(n_{pm})$, i.e., $f: n_{pm} \to l_p$. 

Therefore, for each event $e$, the overload detector calls the event processing latency function $f(n_{pm})$ that gives the estimated event processing latency $l_p$ depending on current number of partial matches in the operator (cf. Algorithm \ref{alg:overloadDetector}, line 3). Using $l_p$ and $l_q$, the overload detector can now compute the estimated event latency $l_e$ (cf. Algorithm \ref{alg:overloadDetector}, line 4).
To build the function $f(n_{pm})$, during run-time, we gather statistics from the operator on the event processing latency $l_p$ for different numbers of partial matches $n_{pm}$. Then, we apply several regression models on these statistics to get the function $f(n_{pm})$, where we  use a regression model that results in lower error. 

We consider the load shedding latency $l_s$ in the inequality (\ref{eq:drop-condition}) since during load shedding no events are processed and hence the event queuing latency is increased by the time needed to drop PMs, i.e., by the load shedding latency $l_s$.  
Similar to the estimated event processing latency, the load shedding latency  $l_s$  also depends on the current number of PMs $n_{pm}$. This is because, the load shedder must sort all current PMs in the operator to find those PMs that have  the lowest utility values (we will show this later).  Therefore, we also represent $l_s$ as a function of $n_{pm}$: $l_s= g(n_{pm})$ (cf. Algorithm \ref{alg:overloadDetector}, line 3). Similarly, to build the function $g(n_{pm})$, during run-time, we gather statistics from the operator on the load shedding latency $l_s$ for different numbers of PMs $n_{pm}$. Then, we apply several regression models on these statistics  to get the function $g(n_{pm})$, where we  use a regression model that results in lower error.

\textbf{Determining overload amount: } 
As we explained above, if the inequality (\ref{eq:drop-condition}) holds, the overload detector calls the LS to drop PMs to avoid violating $LB$ (cf. Algorithm \ref{alg:overloadDetector}, lines 5-9). The question is-- how many PMs must the LS  drop? To answer this question, we need to understand which latency values in the inequality (\ref{eq:drop-condition}) can be controlled. 
 We cannot reduce the event queuing latency $l_q$ and the load shedding latency $l_s$ but we can reduce the event processing latency $l_p$ by dropping some PMs. Therefore, we represent the new event processing latency as $l'_p$ such that the following condition holds.
 
\begin{equation}
	l'_p  + l_q + l_s = LB.
\end{equation}

From the above condition, $l'_p= LB - l_q - l_s$. Therefore, we have to ensure the new processing latency $l'_p$ by dropping a certain number of PMs (denoted by $\rho$).

To compute $\rho$, we should find the number of PMs (denoted by $n'_{pm}$) that impose a latency of $l'_p$ on the operator when processing an event.
Hence, $n'_{pm}$ is a function of $l'_p$. This function is the inverse function $f^{-1}$ of the event processing latency function $f(n_{pm})$, where $f^{-1}: l'_p \to n'_{pm}$. From the inverse function  $f^{-1}$, we can compute the number of PMs $n'_{pm}$.
Keeping only $n'_{pm}$ PMs in the operator's internal state ensures that the operator needs only $l'_p$ time to process an event and hence it maintains $LB$. 
Therefore, the number of PMs to drop $\rho= n_{pm} - n'_{pm}$. For each input event, the overload detector calls the LS to drop $\rho$ partial matches whenever the inequality (\ref{eq:drop-condition}) holds (cf. Algorithm \ref{alg:overloadDetector}, line 9).

Please note that the inequality (\ref{eq:drop-condition}) ensures to keep the event latency $l_e$ less than or equal to $LB$. However, in case of sudden increase in the input event rate or inaccuracy in the functions that predict $l_p$ and $l_s$, there might be a risk of violating $LB$. Therefore, in latency critical applications where $LB$ is a hard bound, we propose to add a safety buffer (denoted by $b_s$) to the inequality (\ref{eq:drop-condition}) as follows:  

\begin{equation}
l_e + l_s + b_s > LB
\label{eq:drop-condition-buffer}
\end{equation}

\begin{algorithm}
	\setbox0\vbox{\small
		{\fontsize{8.0}{9.0}\selectfont
			\begin{algorithmic}[1]
				\algsetblockdefx[function]{func}{endfunc}{}{0.2cm}[3]{#1 \textbf{#2} (#3) \textbf{begin}}{\textbf{end function}}
				
				\func {}{detectOverload}{event e}
				\State $l_q= \mathtt{currentTime()} - \mathtt{e.arrivalTime()}$
				\State $l_p= f(n_{mp})$,  $l_s= g(n_{mp})$ \Comment $n_{mp}$: Current number of PMs.
				\State $l_e= l_q + l_p$
				\If {$l_e + l_s > LB$} \Comment $LB$ might be violated $=>$ drop PMs.
					\State $l'_p= LB- l_q - l_s$
					\State $n'_{pm}= f^{-1}(l'_p)$
					\State $\rho = n_{pm} - n'_{pm}$
					\State $\mathtt{\mathbf{LS}.drop (}\rho)$ \Comment Call LS to drop $\rho$ PMs.
				\EndIf
				\endfunc
			\end{algorithmic}
		}
	}
	\centerline{\fbox{\box0}}
	\caption{Detecting and Determining Overload.}
	\label{alg:overloadDetector}
\end{algorithm}

\subsection{Load Shedding}
In this section, we discuss the functionality of the LS component that is called by the overload detector to drop PMs. The LS  drops PMs with the lowest utility values, where the utility of PMs are learned and stored in $UT$ as we explained in Section \ref{sec:utility}. Algorithm \ref{alg:MarkovDropper} formally explains the functionality of the LS.

%
%

Whenever the LS is called by the overload detector to drop $\rho$ PMs, it needs to know the current $\rho$ PMs in the operator that have the lowest utility values. To get the utility of PMs, the LS simply uses the utility tables given by the model builder. For a PM $pm \subset q_x$ in a window $w$, the LS obtains the utility of $pm$, i.e.,  $U_{pm}$, by a simple lookup in the utility table $UT_{q_x}$. $U_{pm}= UT_{q_x}(i, j)$, where $S_{pm}= s_i \in \mathbb{S}_{q_x}$, and $j= R_w$, i.e., the expected number of events left in the window w (cf. Algorithm \ref{alg:MarkovDropper}, lines 2-4). Therefore, the time complexity to get the utility of a PM is O(1) and hence to get the utility for all current PMs in the operator is O($n_{pm}$).
To find the $\rho$ PMs with the lowest utility values among all PMs, the LS should sort the PMs using their utility values, where a good sorting algorithm can achieve  O($n_{pm}\ log_2 (n_{pm})$) average time complexity (cf. Algorithm \ref{alg:MarkovDropper}, line 5).
After sorting PMs, the LS drops the first $\rho$ PMs which have the lowest utilities, where the LS iterates over the sorted PMs and asks the operator to remove those PMs from its internal state (cf. Algorithm \ref{alg:MarkovDropper}, lines 6-10). This has a time complexity of O($\rho$). Hence, the overall time complexity for the load shedding is O($n_{pm} +  n_{pm}\ log_2 (n_{pm}) + \rho$). As we will show in section \ref{sec:results}, the overhead of our LS is extremely low.

\begin{algorithm}
	\setbox0\vbox{\small
		{\fontsize{8.0}{9.0}\selectfont
			\begin{algorithmic}[1]
				\algsetblockdefx[function]{func}{endfunc}{}{0.2cm}[3]{#1 \textbf{#2} (#3) \textbf{begin}}{\textbf{end function}}
				
				\func {}{drop}{$\rho$}
				
				\LeftComment get utilities of PMs and sort them.
				\For {\textbf{each} {$\mathit{pm} ~ in ~ \mathtt{operator.getPMs()}$}}
					\State $U_{pm}= \mathtt{getUtility(} q_x, S_{pm}, R_w)$, where $pm \subset q_x$
					\State $pmArray.insert(pm)$
				\EndFor
				\State $\mathtt{sortByUtility (}pmArray)$
			
				\LeftComment drop $\rho$ partial matches.
				\For {$index=0  \to \rho$ }
						\If {$index>= pmArray.size()$} \Comment No more PMs to drop!
							\State $\mathbf{return}$
						\EndIf	
						\State $pm= pmArray(index)$
						\State $\mathtt{operator.remove(}pm)$
				\EndFor
			
				\endfunc
			\end{algorithmic}
		}
	}
	\centerline{\fbox{\box0}}
	\caption{Load Shedding.}
	\label{alg:MarkovDropper}
\end{algorithm}

\section{Performance Evaluations}
\label{sec:results}

In this section, we show the performance of \framework~ by evaluating it with three real world datasets and several representative queries.
\subsection{Experimental Setup}

\textbf{\textit{Evaluation Platform.}}
We run our evaluation on a machine which is equipped with 8 CPU cores (Intel 1.6 GHz) and a main memory of 24 GB. The  OS used is CentOS 6.4. We run a CEP operator in a single thread on this machine, where this single thread is used  as a resource limitation. Please note, the resource limitation can be any number of threads/cores and the behavior of pSPICE does not depend on a specific limitation.
We implemented \framework~by extending a prototype CEP framework which is implemented using Java.

\textbf{\textit{Baseline.}}
We also implemented two other load shedding strategies to use as baselines. 1) We implemented a random partial match dropper (denoted by PM-BL) that uses Bernoulli distribution to drop PMs. 2) We also implemented a load shedding strategy (denoted by E-BL) similar to the one proposed in \cite{3:He2014OnLS}.  In addition,  it captures the notion of weighted sampling techniques in stream processing \cite{3:Tatbul:2003:LSD:1315451.1315479}. E-BL drops events from incoming windows, where an event type (e.g., player Id or stock symbol) receives a higher utility proportional to its repetition in patterns and in windows. 
Then, depending on event type utilities, it uses uniform sampling to decide which events to drop from the same event type.

\textbf{\textit{Datasets.}}
We use three real-world datasets. 
1) A stock quote stream from the New York Stock Exchange, which contains real intra-day quotes of 500 different stocks from NYSE collected over two months from Google Finance \cite{google_finance}.
2) A position data stream from a real-time locating system (denoted by RTLS) in a soccer game \cite{1:debs2013}. Players, balls, and referees  are equipped with sensors that generate events containing their position, velocity, etc. 
3) Public bus traffic (denoted by PLBT) from a real transportation system in Dublin city \cite{Zacheilas:2015}. It contains events from 911 buses, where each event has several information about those buses, e.g., location, stop, delayed, etc. 

\textbf{\textit{Queries.}}
We apply four queries (Q1, Q2, Q3, Q4) that cover an important set of operators in CEP: sequence operator, sequence operator with repetition, sequence with any operator, and any operator, all with skip-till-next/any-match \cite{1:Balkesen:2013:RRI:2488222.2488257, 1:cu2010tesla, Dayarathna:2018:RAE:3186333.3170432, Wu:2006:HCE:SASE}. Moreover, the queries use both \textbf{time-based} and \textbf{count-based} sliding window strategies with \textbf{different predicates}. 
The queries are as follows:

Q1 (sequence operator): detects a complex event  when rising or falling stock quotes of 10 certain stock symbols (defined as $RE$ or $FE$, respectively) are detected within $\mathit{ws}$ events in a certain sequence.  Q1 is of form: seq ($RE_1$; $RE_2$;..;$RE_{10}$) or seq ($FE_1$; $FE_2$;..;$FE_{10}$), where $RE_x$ or $FE_x$ is  rising/falling event of the stock company $x$.

Q2 (sequence operator with repetition): detects a complex event when 10 rising or 10 falling stock quotes of certain stock symbols (defined as $RE$ or $FE$, respectively) \textit{with repetition} are detected within $\mathit{ws}$ events in a certain sequence. Q2 is of form: seq ($RE_1$;  $RE_1$; $RE_2$; $RE_3$; $RE_2$; $RE_4$; $RE_2$; $RE_5$; $RE_6$; $RE_7$; $RE_2$; $RE_8$; $RE_9$; $RE_{10}$), where $RE_x$ is defined as in Q1. The sequence for falling quotes is similar. 

Q3 (sequence with any operator): uses the RTLS dataset. It detects a complex event when any $n$ defenders of a team (defined as $\mathtt{DF}$) defend against a striker (defined as $\mathtt{STR}$) from the other team within $\mathit{ws}$ seconds from the ball possessing event by the striker. The defending action is defined by a certain distance between the striker and the defenders. We use two players as strikers; one striker from each team. Q1 is of form: seq ($STR$; any ($n$, $DF_1$, $DF_2$, .., $DF_n$)), where $DF_x$ is the defend event of the player $x$.

Q4 (any operator): uses the PLBT dataset. It detects a complex event when any $n$ buses  within a window of size $ws$ events get delayed at the same stop. Q4 is of form: any ($B_1$, $B_2$, ..., , $B_n$), where $B_x$  is the bus event.

\subsection{Experimental Results}
\label{sec:exp_results}
In this section, we evaluate the performance of \framework. First, we show its impact on QoR, i.e., number of false negatives, and compare it with PM-BL and E-BL. Then, we show the importance of using the processing time of a PM in calculating its utility. Finally, we present the overhead of \framework.

If not stated otherwise, we use the following settings. 
For Q1 and Q2, we use a \textit{count-based} sliding window. For both queries, we use  a \textit{logical} predicate where a new window is opened for each incoming event of the leading stock symbols. We choose 4 important companies as leading stock companies. Q3 uses a \textit{time-based} sliding window. Again, we use  a \textit{logical} predicate for Q3, where a new window is opened for each incoming striker event (STR).
For Q4, we use  a \textit{count-based} sliding window and a \textit{count-based} predicate, where a new window is opened every 500 events, i.e., slide size is 500 events.
We stream events to the operator from datasets that are stored in files where we first stream events at event input rates which are less or  equal to the maximum operator throughput until the model is built. After that, we increase the input event rate to enforce load shedding as we will mention in the following experiments.  
The used latency bound $LB = 1$ second. 
We execute several runs for each experiment and show the mean value and standard deviation.

\textbf{\textit{Impact on QoR and the given latency bound.}}
\label{sec:quality-fn}
Now, we show the performance of \framework~w.r.t. its impact on QoR (i.e., number of false negatives) and maintaining the given latency bound.  Two factors influence the performance of \framework: 1) match probability, and 2) input event rate.  Match probability represents the percentage  of PMs that complete and become complex events out of all PMs. It is computed from the ground-truth by dividing the total number of complex events by the total number of PMs. We can control the match probability by varying the pattern size and/or the window size.

\paragraph{Impact of match probability} 
\label{sec:match_probability}
To evaluate the performance of \framework~with different match probabilities, we run experiments with Q1, Q2, Q3 and Q4. For Q1 and Q2, we use a variable  window size to control the match probability since Q1 and Q2 have a fixed pattern size. Higher is the window size, higher is the match probability. 
We use the following window sizes for Q1: $ws= $ 3.5K,  4.5K, 5K, 5.5K, 6K, 10K events. For Q2, the used window sizes are: $ws= $ 6K, 7K, 7.5K, 8K, 12K, 14K events. For Q3 and Q4, we use a fixed window size but a variable pattern size. For Q3, we use a window size $ws$ of  15 seconds and the following pattern sizes (i.e., number of defenders):  $n$= 2, 3, 4, 5, 6. The window size $ws$ for Q4 is 8K events and we use the following pattern sizes (i.e., number of buses): $n= $ 3, 4,  7, 8, 10. Moreover, we stream all datasets to the operator with an event input rate that is higher than the maximum operator throughput by 20\% (i.e.,  event rate= 120\% of the maximum operator throughput).

\begin{figure*}[t]
	\centering
	\begin{subfigure}[t]{0.22\linewidth}
		\includegraphics[width=0.99\linewidth]{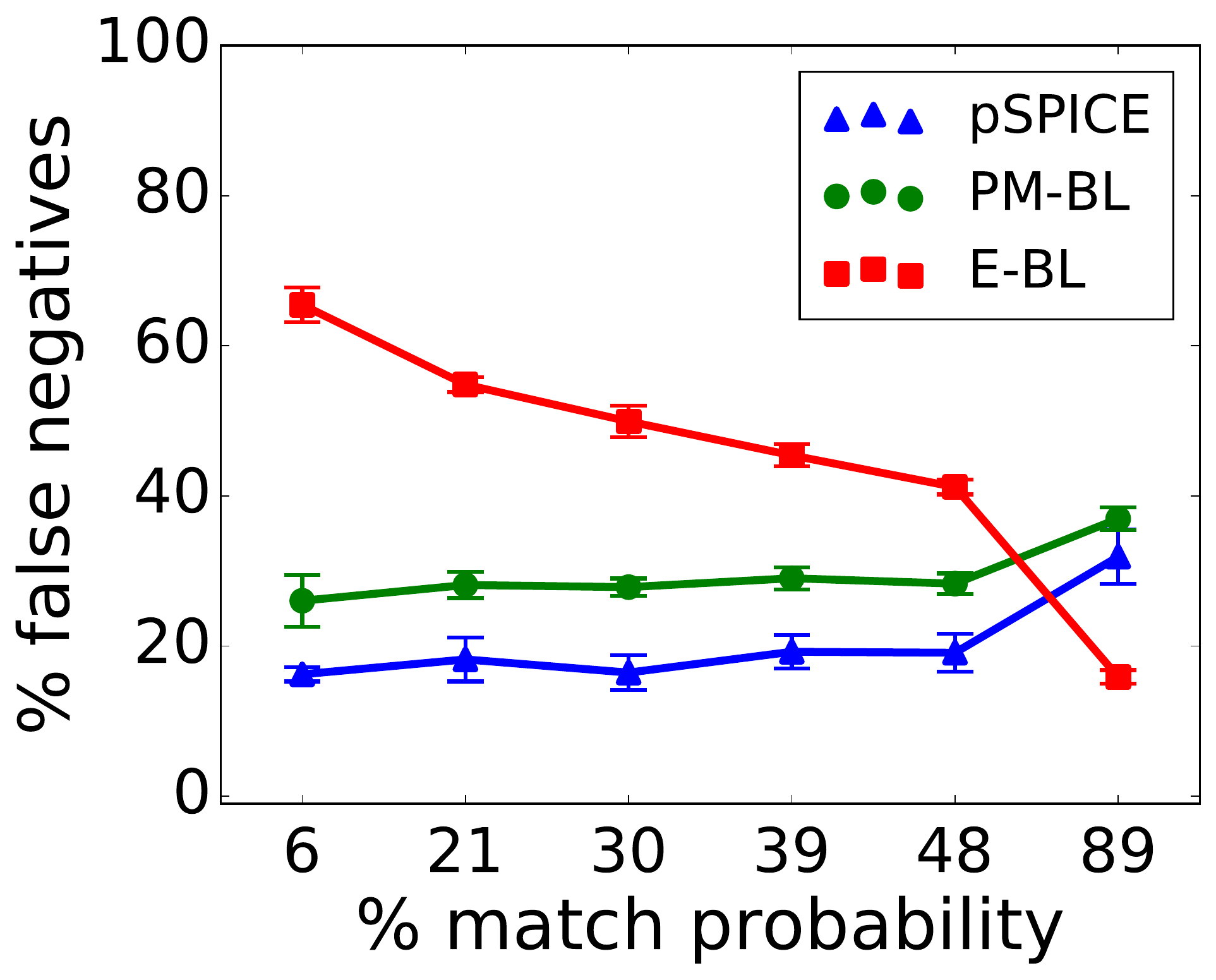}
		\caption[]{Q1}
		\label{fig:quality-q1-r1}
	\end{subfigure}
	\hfil%
	\begin{subfigure}[t]{0.22\linewidth}
		\includegraphics[width=0.99\linewidth]{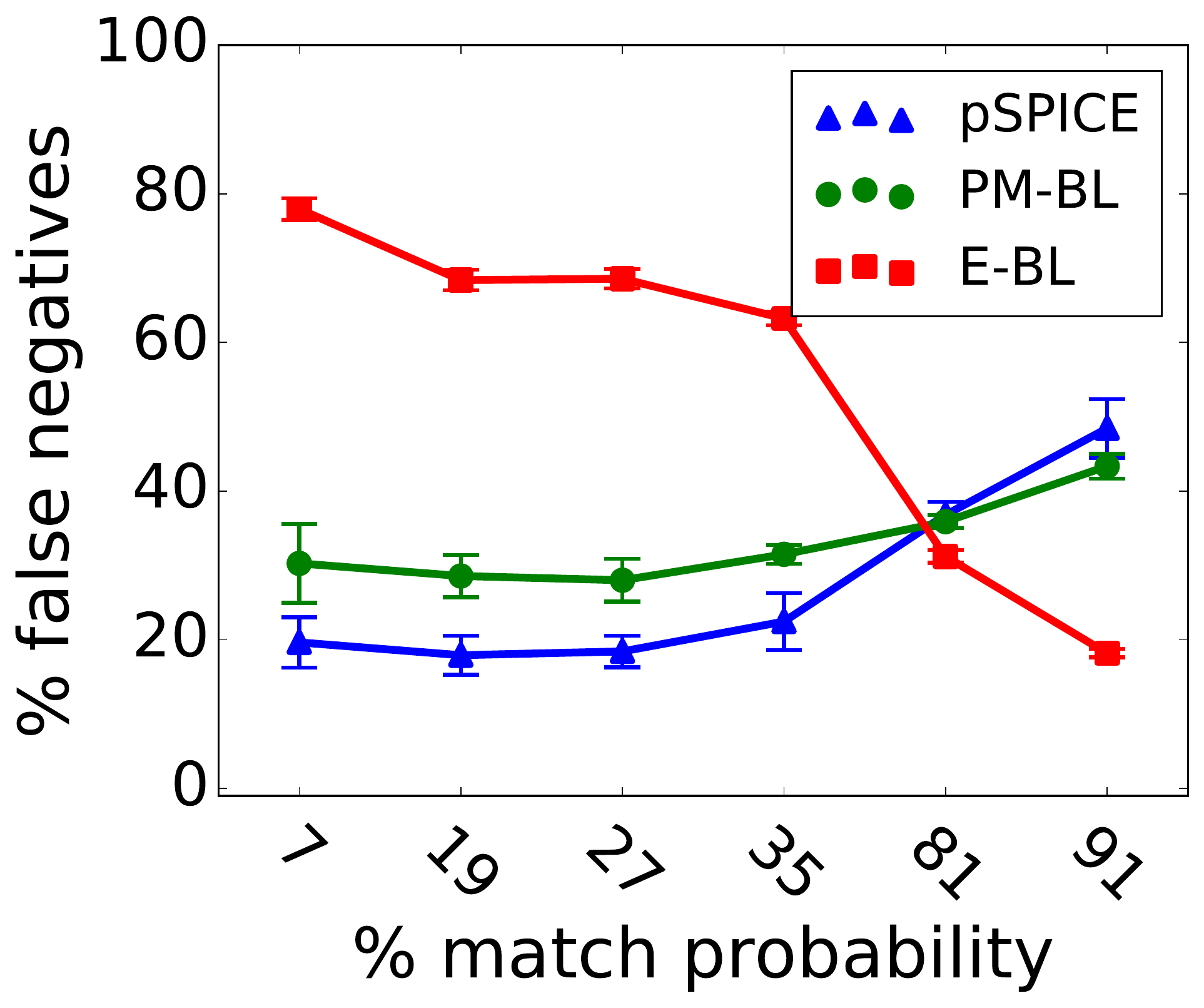}
		\caption[]{Q2}
		\label{fig:quality-q2-r1}
	\end{subfigure}
	\hfil%
	\begin{subfigure}[t]{0.22\linewidth}
		\includegraphics [width=0.99\linewidth]{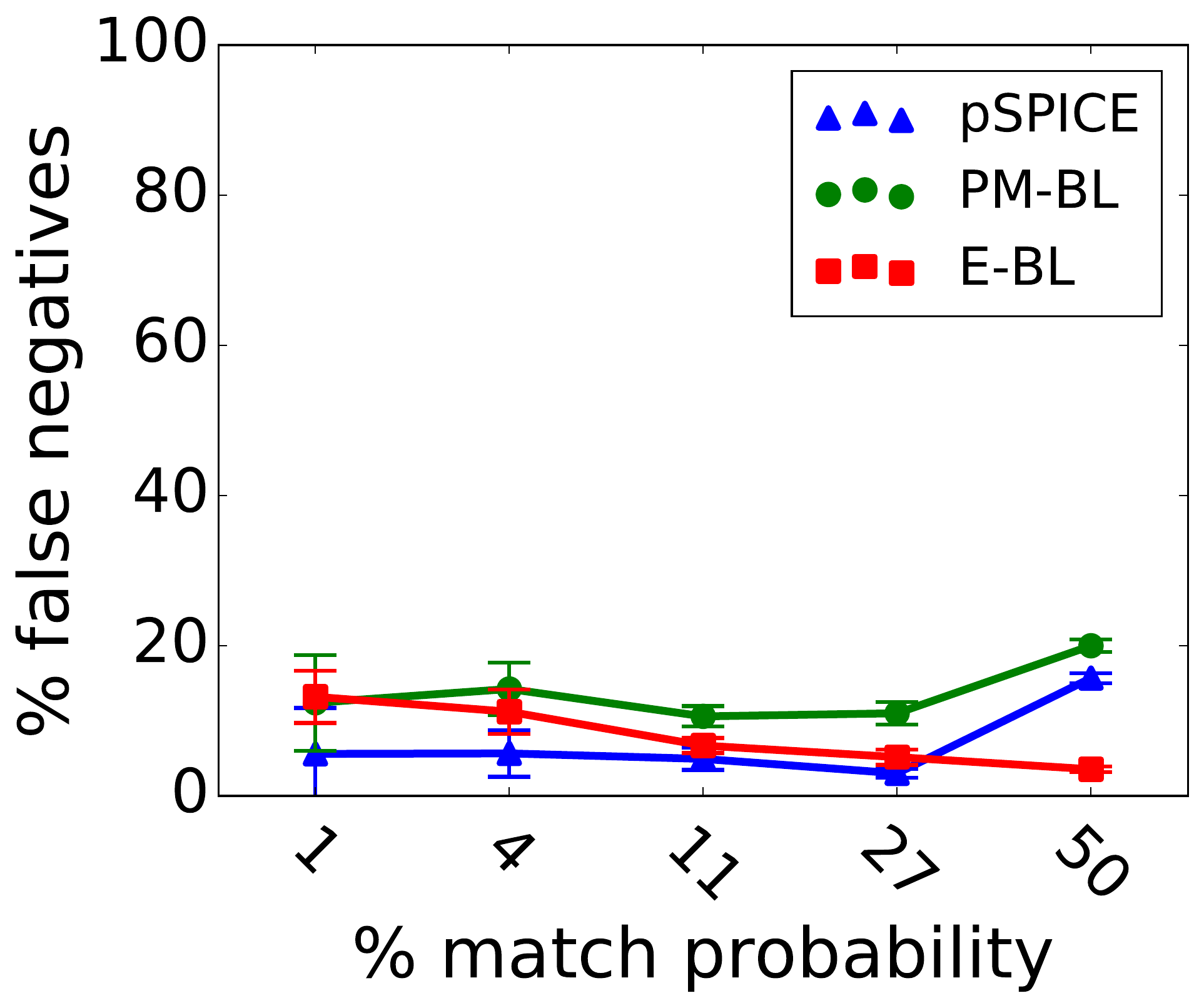}
		\caption[]{Q3}
		\label{fig:quality-q3-r1}
	\end{subfigure}
	\hfil%
	\begin{subfigure}[t]{0.22\linewidth}
		\includegraphics [width=0.99\linewidth]{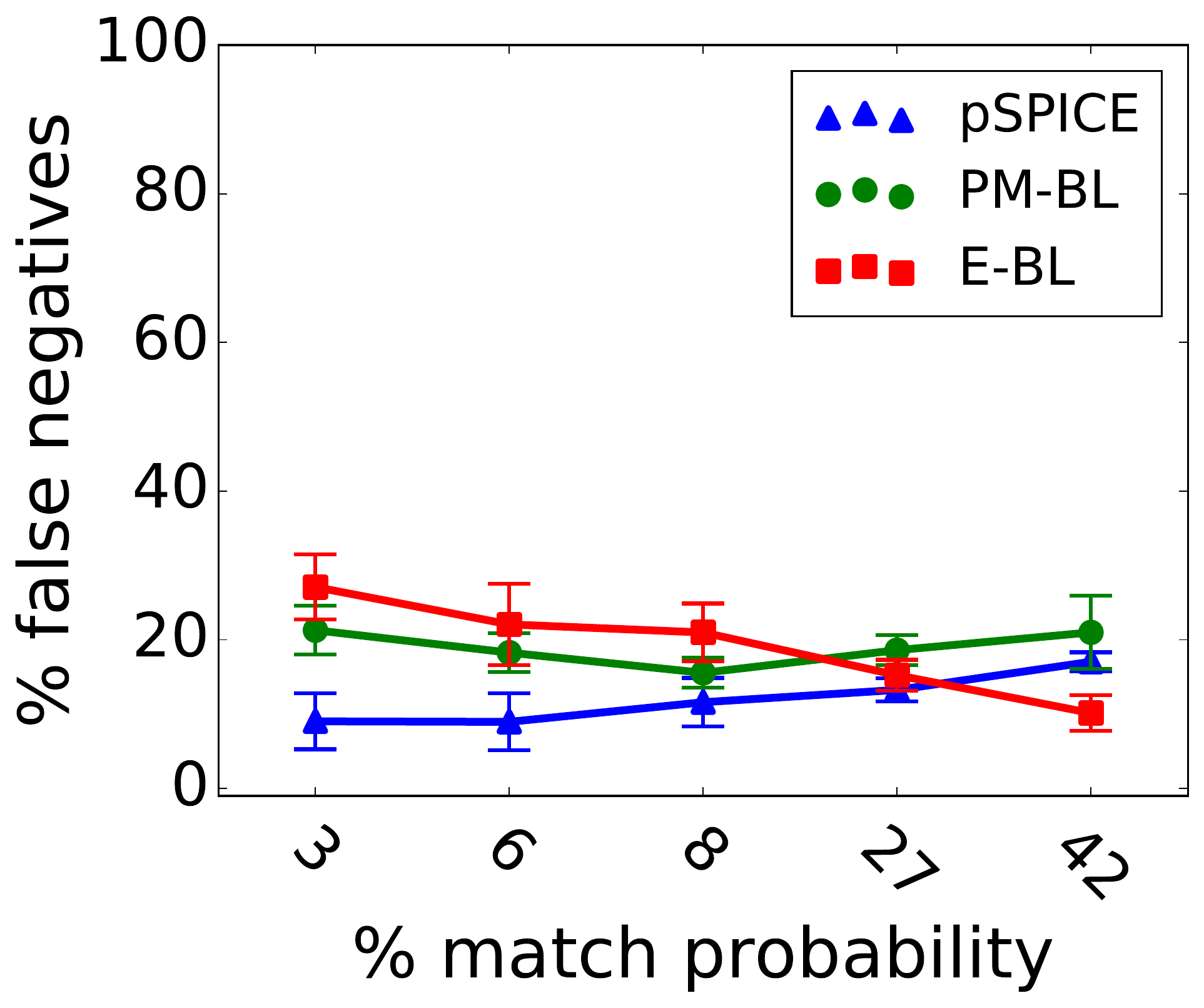}
		\caption[]{Q4}
		\label{fig:quality-q4-r1}
	\end{subfigure}
	\caption{Impact of match probability.}
	\label{fig:quality-r1}
\end{figure*}

Figure \ref{fig:quality-r1} shows results for all queries, where the x-axis represents the match probability and the y-axis represents the percentage of false negatives.
A low match probability means that most of the PMs don't complete and hence dropping those PMs that will not complete decreases the dropping impact on QoR. On the other hand, a high match probability means that most of the PMs complete and become complex events and hence dropping any PM may result in a false negative. This is observed in Figure \ref{fig:quality-r1} for all queries (Q1, Q2, Q3, Q4).
Figure \ref{fig:quality-q1-r1} depicts the results for Q1, where it shows that the percentage of false negatives produced by \framework~increases with increasing match probability. It increases from 16\% to 32\% when the match probability increases from 6\% to 89\%, respectively. We observed a similar behavior for PM-BL, where the percentage of false negatives increases from 26\% to 37\% when the match probability increases from 6\% to 89\%, respectively.
As we observe from the figure, a high match probability degrades the performance of \framework~since dropping any PM might result in a false negative as all PMs have similar completion probability.
In this experiment, \framework~reduces the percentage of false negatives by up to 70\% compared to PM-BL.
Please note that a high rate of PM drop is because the operator load doesn't come only from processing PMs but also from managing windows and events and checking whether an event opens a partial match. 

The performance of E-BL is bad when the match probability is low and it becomes better with higher match probability as shown in Figure \ref{fig:quality-q1-r1}. This is because, a low match probability means a small window size where the probability to drop an event that matches the pattern is high and the probability to find an event as a replacement for the dropped event to match the pattern is low. On the other hand, with a higher match probability (i.e., a larger window size), the probability to drop an event that matches the pattern is low and the probability to find an event as a replacement for the dropped event to match the pattern is high. Hence, the percentage of false negatives  decreases with a higher match probability.
In the figure, the percentage of false negatives, for E-BL, is 65\% and 16\% when the match probability is 6\% and 89\%, respectively.
\framework~reduces the percentage of false negatives by up to 300\% compared to E-BL when the match probability is not too high.
For a high match probability (cf. Figure \ref{fig:quality-q1-r1}, in case match probability is 89\%),  E-BL outperforms \framework~.
However, please note, in CEP, it is unrealistic to have such a high match probability that implies completion of most PMs.

Figure \ref{fig:quality-q2-r1}, using Q2, shows similar behavior to the results of Q1. The percentage of false negatives for \framework~and PM-BL increases again with increasing match probability. However, \framework~results in a lower percentage of false negatives by up to 58\% compared to PM-BL till 81\% match probability. After that, PM-BL outperforms \framework. This is because, as we mentioned above, all PMs have a high probability to complete and become complex events and hence it is hard for \framework~to decide which PM to drop. Besides that, \framework~has a slightly higher overhead than PM-BL which results in dropping more PMs and hence resulting in more false negatives. The results for E-BL is similar to the results in Q1.


In Figure \ref{fig:quality-q3-r1}, using Q3, the percentage of false negatives produced by \framework~and PM-BL also increases with increasing the match probability. \framework~results in reducing the percentage of false negatives by up to 92\% compared to PM-BL. As in Q1 and Q2, E-BL produces less false negatives when the match probability increases.  A higher match probability in Q3 means a smaller pattern size (in the figure, the match probability 50\% corresponds to a pattern of size $n= $ 2) which makes it easy to find a replacement event to match the pattern instead of a dropped event. The results for Q3, compared to the results for Q1 and Q2, show  that E-BL outperforms \framework~with a smaller match probability (after 27\%). This is because Q3 uses \emph{any} operator which means any event can match the pattern. Hence, the probability to find  a replacement for a dropped event is much higher in Q3 compared to Q1 and Q2 which matches a \emph{sequence} of certain event types (stock symbol/company). Please note that, in Q1 and Q2, only the same event type can replace a dropped event of that type.
Figure \ref{fig:quality-q4-r1}, using Q4, shows similar results to the results of Q3 since the query of bus data is similar to the query of soccer data (i.e., Q3). As a result, we skip explaining it.

\begin{figure}[t]
	\centering
	\begin{subfigure}[t]{0.45\linewidth}
		\includegraphics[width=0.99\linewidth]{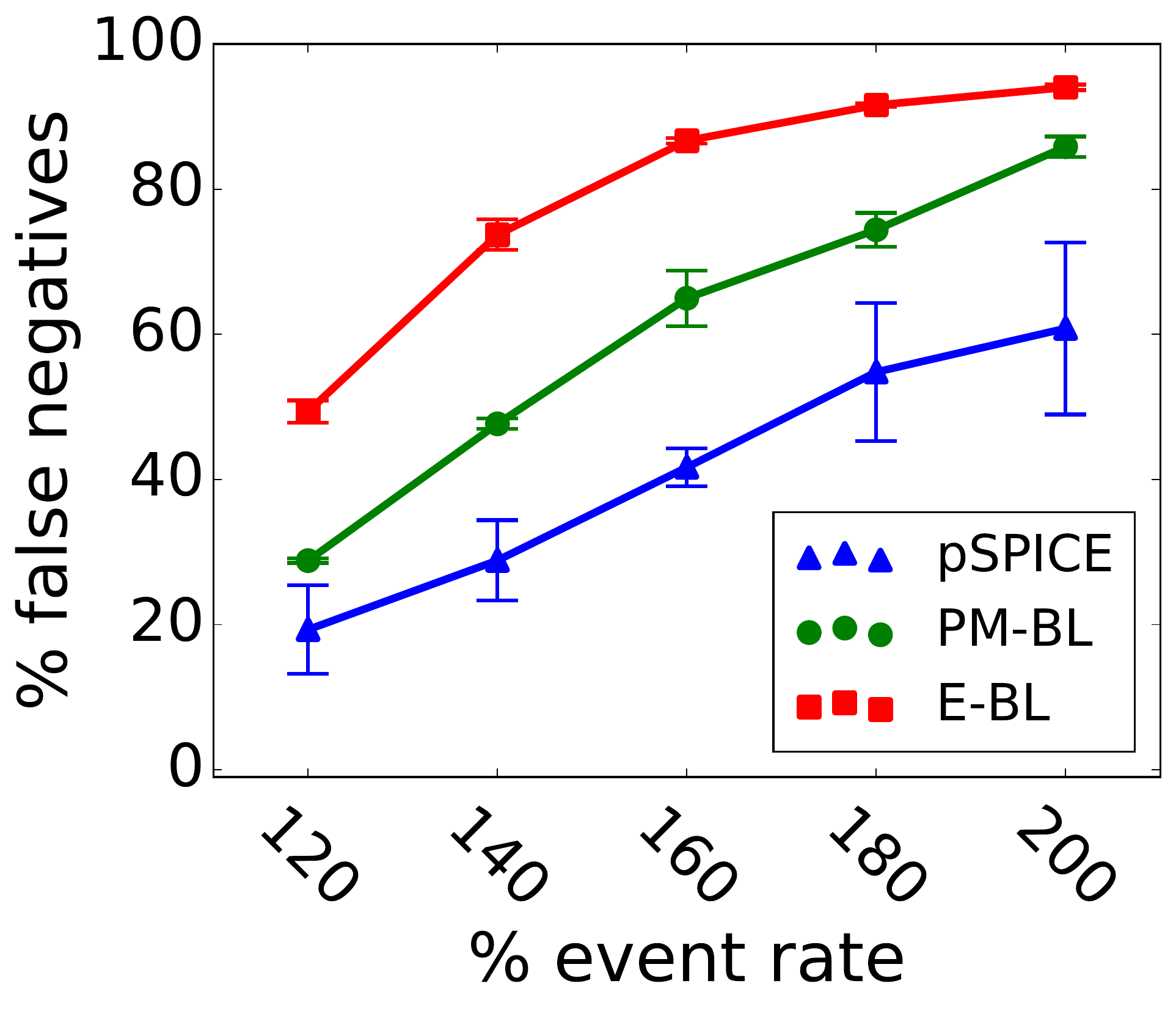}
		\caption[]{Q1}
		\label{fig:eventRate-q1}
	\end{subfigure}
	\hfil%
	\begin{subfigure}[t]{0.45\linewidth}
		\includegraphics[width=0.99\linewidth]{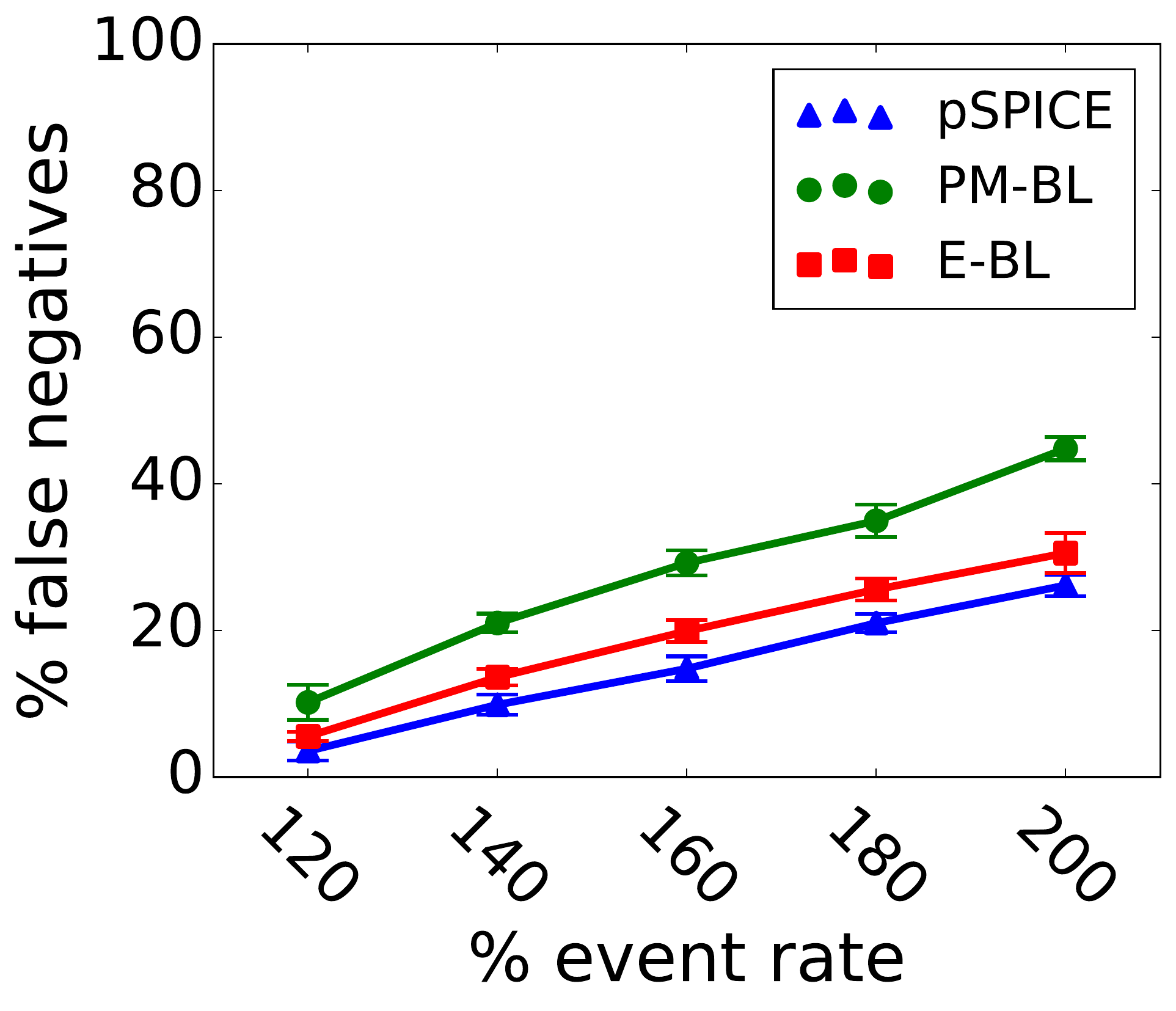}
		\caption[]{Q3}
		\label{fig:eventRate-q3}
	\end{subfigure}
	\caption{Impact of event rate.}
	\label{fig:eventRate}
\end{figure}

\paragraph{Impact of event rate}
To evaluate the impact of input event rate on the performance of \framework, we run experiments with Q1, Q2, Q3, and Q4 using the same setting as in the above section (cf. Section \ref{sec:exp_results}-a). However, to show the impact of different event rates, we  streamed all datasets to the operator with  event input rates that are higher than the maximum operator throughput by 20\%, 40\%, 60\%, 80\%, and 100\%  (i.e.,  event rate= 120\%, 140\%, 160\%, 180\%, 200\%, of the maximum operator throughput). In addition, we used a fixed match probability for all queries. Figure \ref{fig:eventRate} depicts the impact of input event rates for Q1 and Q3, where the x-axis represents the event rate and the y-axis represents the percentage of false negatives. We use a match probability of 30\% for Q1 and 4\% for Q3. The results for Q2 and Q4 show similar behavior, hence we don't show them. 

It is clear that using a higher event rate results in dropping more partial matches and hence increasing the percentage of false negatives. In Figure \ref{fig:eventRate-q1}, using Q1, the percentage of false negatives for \framework~increases with increasing the event rate, where it is 18.5\% and 60\% when the even rate is 120\% and 200\%, respectively. The same behavior is observed for PM-BL and E-BL. The percentage of false negatives for PM-BL increases from 29\% to 86\% and for E-BL from 49\% to 94\%, with the two event rates. 
Please note that  for the considered match probability \framework~is consistently better than PM-BL and E-BL irrespective of the event rate.
Figure \ref{fig:eventRate-q3}, using Q3, as expected, shows similar behavior.

\begin{figure}[t]
	\begin{minipage}[t]{0.45\linewidth}
		\includegraphics[width=0.99\linewidth]{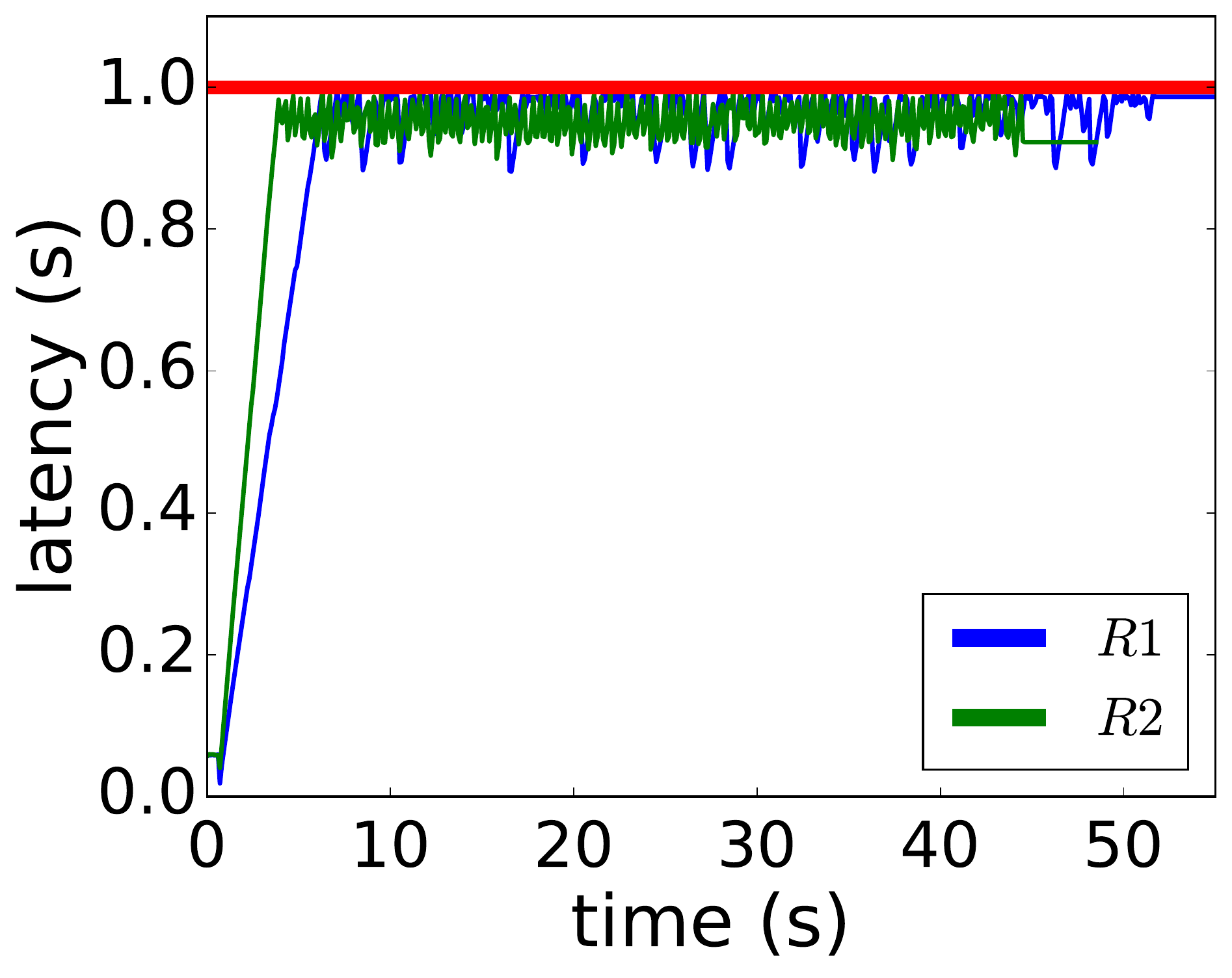}
		\caption{event latency $l_e$}
		\label{fig:lb}	
	\end{minipage}
		\hfil%
	\begin{minipage}[t]{0.45\linewidth}
		\includegraphics[width=0.99\linewidth]{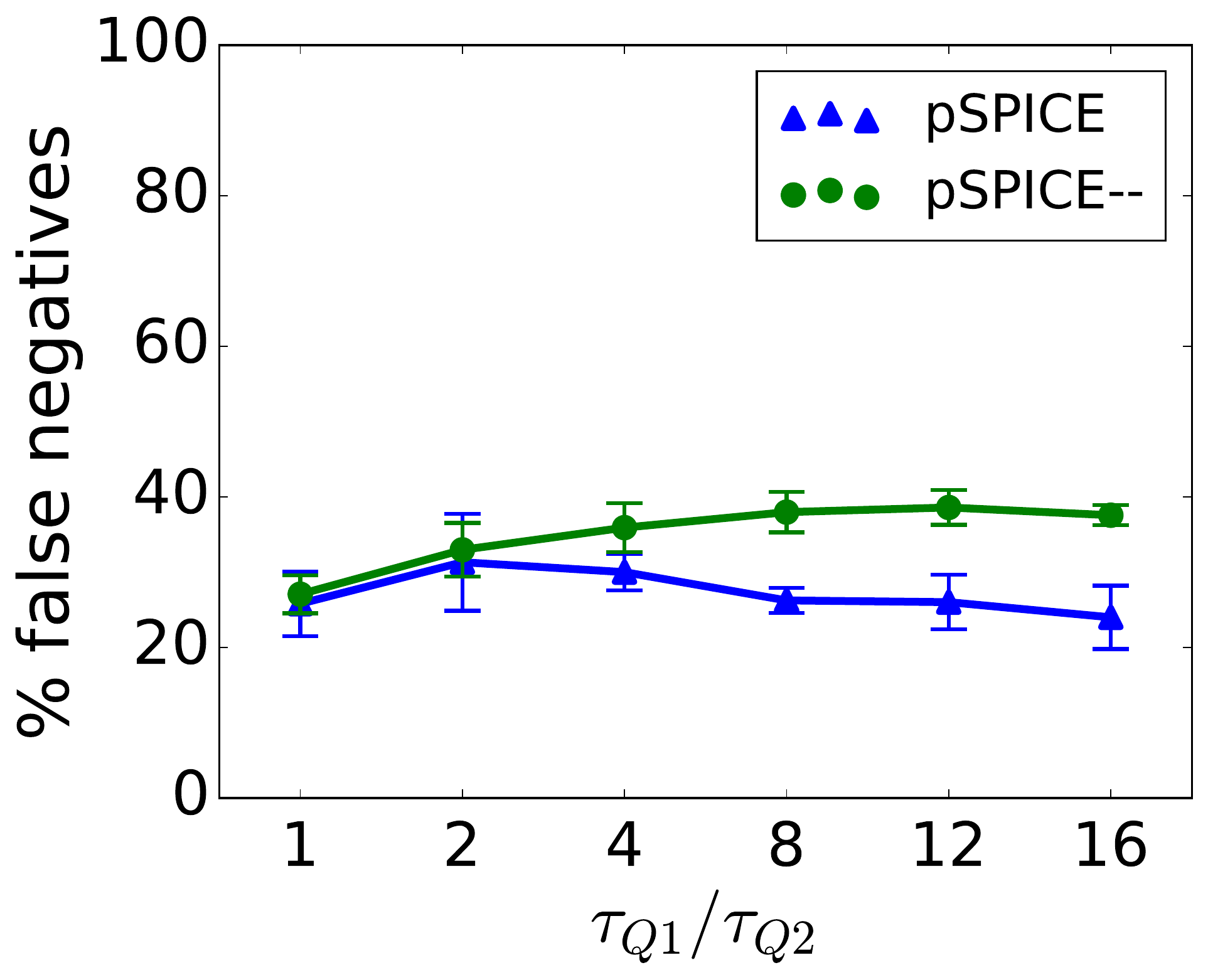}
		\caption{processing time $\tau_{pm}$}
		\label{fig:processing-time}	
	\end{minipage}
\end{figure}

\paragraph{Maintaining $LB$}
\framework~performs load shedding to maintain a given latency bound. Figure \ref{fig:lb} shows the result for running Q2 with two event rates 120\% (defined as $R1$) and 140\% (defined as $R2$). In the figure, the x-axis represents time and the y-axis represents the event latency $l_e$. We observed similar results for other event rates and queries and hence we don't show them.
The figure shows that \framework~always maintains the given latency bound $LB$ which is 1 second in this experiment, regardless of the event rate.

\textbf{\textit{Impact of processing time of a PM ($\tau_{pm}$) on utility  calculation.}}

As mentioned above, the completion probability $P_{pm}$ of a partial match $pm$ is a good indicator to know  whether  $pm$ will complete or not, and therefore, we use it in calculating the utility of PMs (cf. Equation (\ref{eq:ut})). However, the processing time of a PM ($\tau_{pm}$) is also an important factor in calculating the utility of a PM, and therefore, we use it in deriving the utility of PMs as well (cf. Equation (\ref{eq:ut})). To support this argument, we run experiments using \framework~with two different ways of calculating the utility of PMs as follows: 1) using Equation (\ref{eq:ut}) where we consider both the completion probability and processing time of PMs in calculating the utility of PMs and 2) considering only the completion probability in calculating the utility of PMs (i.e., the denominator in Equation (\ref{eq:ut}) is 1). We refer to the load shedding strategy that considers only the completion probability  in calculating the utility of PMs as \framework-\,-.

To evaluate the performance of \framework~and \framework-\,-, we run both Q1 and Q2 in the same operator and use a window of size 10K and a pattern weight of one for both queries. The used event rate is 120\%. Since we intend to analyze the impact of processing time in calculating the utility of PMs on QoR, we force the processing time of Q1 to be higher than the processing time of Q2 by a factor. We refer to this factor as $\tau_{Q1}/\tau_{Q2}$, where we use the following values: $\tau_{Q1}/\tau_{Q2}= $ 1, 2, 4, 8, 12, 16. Figure \ref{fig:processing-time} depicts the percentage of false negatives for \framework~and \framework-\,-. In the figure, the x-axis represents the factor $\tau_{Q1}/\tau_{Q2}$ while the y-axis represents the percentage of false negatives.

In the figure, the performance of \framework~and \framework-\,- is same for low factors $\tau_{Q1}/\tau_{Q2}$. This is because the processing time of PMs in Q1 and Q2 have less impact on the utility. The difference between the percentage of false negatives between \framework~and \framework-\,- increases when the factor $\tau_{Q1}/\tau_{Q2}$ increases. The percentage of false negatives for \framework~is 23\% when $\tau_{Q1}/\tau_{Q2}= $ 16 while it is 37.5\% for \framework-\,- with the same factor. 
This shows that \framework~results in reducing the percentage of false negatives by 62\% compared to \framework-\,- for  $\tau_{Q1}/\tau_{Q2}=$ 16.
As a result, we support our claim that considering the processing time of PMs is an important factor in calculating the utility of PMs.

\textbf{\textit{\framework~overhead.}}
Next, we show the overhead of \framework~both during load shedding and during model building.

\begin{figure}[t]
	\centering
	\begin{subfigure}[t]{0.45\linewidth}
		\includegraphics[width=0.99\linewidth]{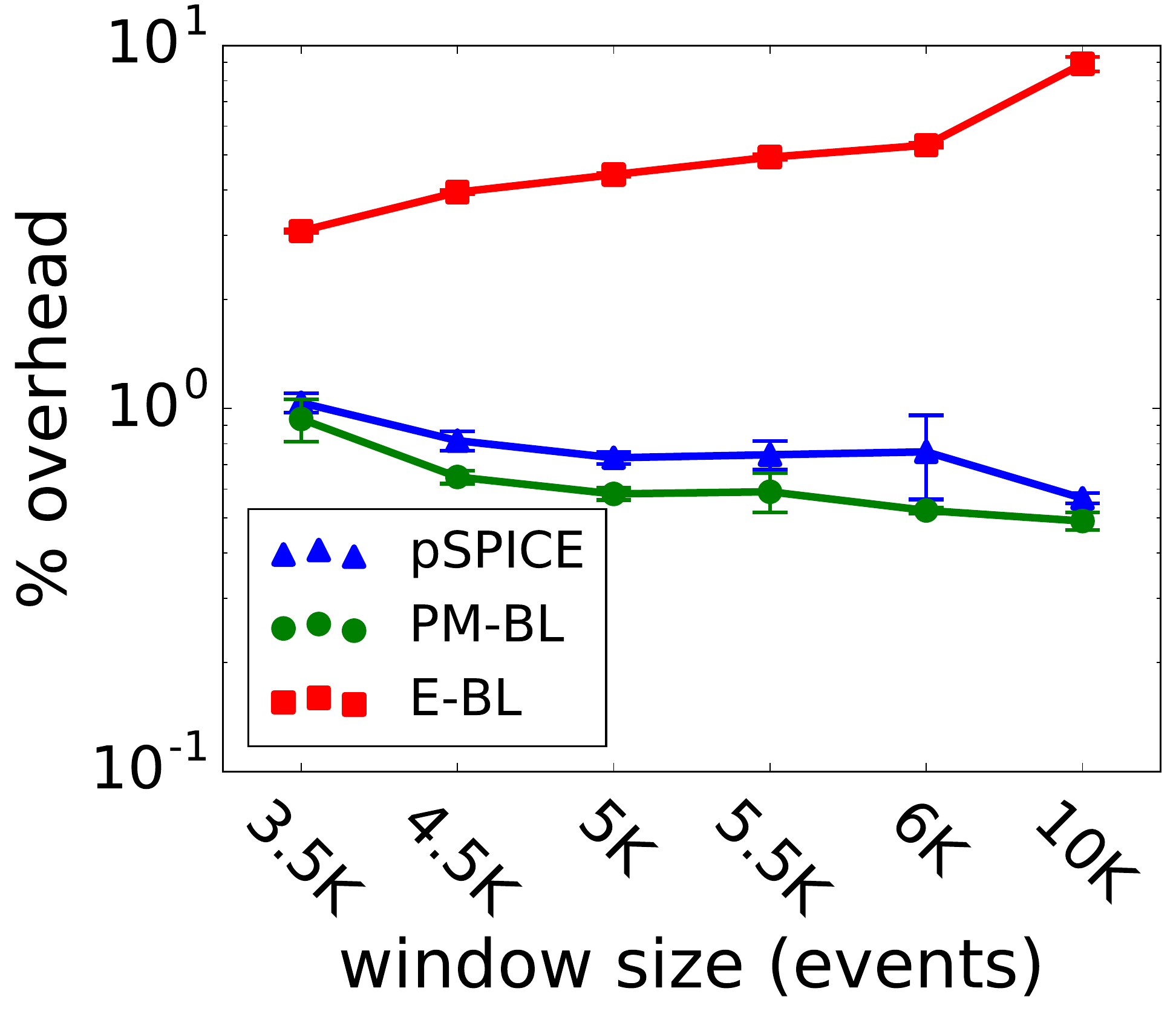}
		\caption[]{load shedding overhead}
		\label{fig:ls-overhead-q1-r1}
	\end{subfigure}
	\hfil%
	\begin{subfigure}[t]{0.45\linewidth}
		\includegraphics[width=0.99\linewidth]{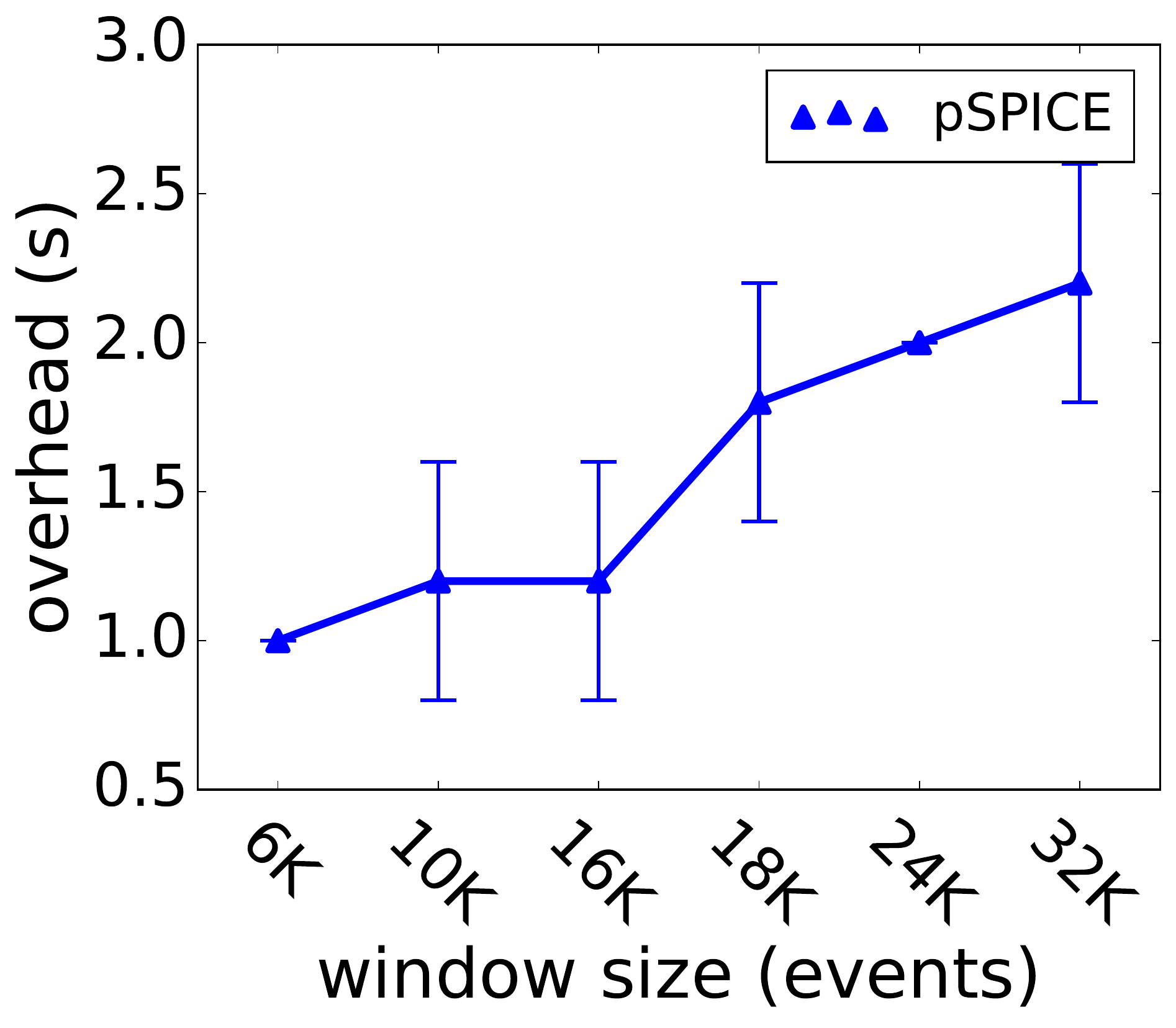}
		\caption[]{model overhead}
		\label{fig:model-overhead-q2-r1}
	\end{subfigure}
	\caption{overhead of \framework.}
	\label{fig:overhead-r1}
\end{figure}

\paragraph{Load shedding overhead}
The load shedder and the overload detector are time-critical tasks and their overhead directly affects QoR, therefore, they must be light-weight. To show the overhead of the load shedder and overload detector components in \framework, we run experiments with all queries using the same setting as in Section (\ref{sec:exp_results}-a). Figure \ref{fig:ls-overhead-q1-r1} depicts the results for Q1, where the x-axis represents the used window size and the y-axis (log scale) represents the percentage of overhead compared to the total time that the operator needs to process the input dataset. We observed similar results for Q2, Q3, and Q4 and hence we don't show them.  

In the figure, the overhead of \framework~is 1\% in case the window size $ws$  is 3.5K. The overhead of \framework~decreases with increasing the window size, where the overhead is 0.7\% when the window size is 10K. This is because a higher window size means that more windows are overlapped. Since events are processed in each window, higher is the window overlap, higher is the processing latency of events and hence lower is the operator throughput. A low operator throughput results in having a smaller load shedding overhead as a percentage value. The overhead of PM-BL is slightly lower than the overhead of \framework~which is expected since PM-BL performs random PMs shedding and doesn't have any cost for sorting PMs.
The overhead of E-BL is 3\% in case of window size of 3.5K and increases with increasing the window size, where it is 10\% with a window size of 10K. The reason is again related to the number of overlapped windows. With high window overlap, since E-BL drops events from windows, it must drop more events in total and hence it causes more overhead. This shows that the overhead of \framework~is lower than the overhead of E-BL by up to 1400\%. As a result, dropping PMs has less overhead than dropping events since it is performed on a higher granularity.

\paragraph{Model overhead}
As we mentioned above, building the model is not a time-critical task. However, since there might be a need to retrain the model in case the distribution of input event stream and/or the content of input events change (cf. Section \ref{sec:retrain}), we also analyze  the overhead of building the model in \framework.
An important factor that controls the overhead of building the model is the window size since it represents the number of iteration in the value iteration algorithm. Higher is the window size, more iterations is needed to solve Markov reward process and hence higher is the overhead.

To evaluate the overhead of building the model, we run experiments with Q1 with the same setting as in Section (\ref{sec:exp_results}-a) but we use higher window sizes to show its impact on the overhead. We use the following window sizes: $ws= $ 6K, 10K, 16K, 18K, 24K, 32K events. Figure \ref{fig:model-overhead-q2-r1} shows the overhead of model building in \framework, where the x-axis represents the window size and the y-axis represents the time needed in seconds. In the figure, as expected, the model building overhead increases with increasing the window size, where it is 1 second when window size is 6K events and 2.4 seconds when window size is 32K events. However, this overhead is still small which means that the model can be retrained without introducing a high overhead on the system or waiting a long time for a new model.

\paragraph{Discussion}
Through extensive evaluations with several datasets and a set of representative queries, \framework~shows that it has a very good performance w.r.t. QoR where it usually outperforms both PM-BL and E-BL, especially with \emph{sequence} operator and \emph{sequence with repetition} operator. Only in the scenario of a relatively high match probability, E-BL might outperform \framework, especially for \emph{any} operator. 
However, E-BL, as mentioned in Section \ref{sec:introduction} and \ref{sec:background}, might result in false positives, e.g., for negation operator.
Moreover, \framework~ reduces the overhead of load shedding significantly compared to E-BL which has a high overhead.  The overhead of \framework~is only slightly higher than the overhead of PM-BL.


\section{Related Work}
\label{sec:related_work}
%
%
%
%
%

CEP is used to detect patterns in input event streams which are continuous and infinite \cite{4812454, Zacheilas:2015, spectre:2017, 1:Balkesen:2013:RRI:2488222.2488257, Wu:2006:HCE:SASE, 8606633}. 
In CEP, there are several well-known operators:  sequence, negation,  disjunction, conjunction, aperiodic, and periodic \cite{4812454, Wu:2006:HCE:SASE, Cadonna:2011:SES:1951365.1951372},  where the order of events in the input event stream and in patterns is extremely important, e.g., for sequence and negation operators. 

The input event streams, in CEP, have high volume and usually need to be processed at near real-time \cite{Quoc:2017:SAC:3135974.3135989, CastroFernandez:2013:ISO:2463676.2465282}. To process those events within a given latency bound, researchers have proposed several techniques such as parallelism, optimizations, and pattern sharing. In \cite{spectre:2017, 1:Balkesen:2013:RRI:2488222.2488257, 1:Zeitler413076, CastroFernandez:2013:ISO:2463676.2465282, Zacheilas:2015, 1:s4:5693297}, the authors proposed to distribute the CEP operator graph on multiple compute nodes and to parallelize each operator on one (scale-up) or more nodes (scale-out). To efficiently process patterns, in \cite{Wu:2006:HCE:SASE, Woods:2010:CED:1920841.1920926}, the authors proposed different optimizations, e.g., intra- and inter-operator optimizations \cite{Wu:2006:HCE:SASE}, or using a special hardware (FPGA) to speedup the event processing \cite{Woods:2010:CED:1920841.1920926}. Another way to improve the operator throughput is by sharing the pattern matching between several patterns as proposed in  \cite{Ray:2016:SPS:2882903.2882947, Schultz-Moller:2009:DCE:1619258.1619264}. The authors proposed algorithms to find the best sharing between different patterns in an operator.

The above mentioned techniques may not always be possible or may not be sufficient to handle the incoming event rate. Therefore, load shedding is used in these situations to avoid violating a defined latency bound. Various approximation techniques are frequently used to avoid resource constraints in various domains such as distributed graph processing~\cite{shang2014auto}, in-network processing~\cite{Bhowmik:Expressive, Bhowmik:Addressing}, stream processing~\cite{3:Tatbul:2003:LSD:1315451.1315479, 3:Rivetti:2016:LSS:2933267.2933311, 3:Olston:2003:AFC:872757.872825}, etc. Load shedding has, especially, been extensively studied in the stream processing domain \cite{3:Tatbul:2003:LSD:1315451.1315479, 3:Tatbul:2006:WLS:1182635.1164196, 3:Rivetti:2016:LSS:2933267.2933311, 3:Olston:2003:AFC:872757.872825, 3:Kalyvianaki:2016:TFF:2882903.2882943, 8622265:concept-drift, Quoc:2017:SAC:3135974.3135989, Tok:2008:SAP:1353343.1353414}.
The main focus here is on individual tuples, where the authors assume that the importance/utility of  tuples are independent from other tuples. 

In \cite{3:Tatbul:2003:LSD:1315451.1315479, 3:Olston:2003:AFC:872757.872825, 8622265:concept-drift}, the authors assume that tuples have the same processing latency but different utilities depending on the tuple's content. Hence, if there is a need to drop tuples, they drop those tuples with the lowest utilities.
\cite{3:Tatbul:2003:LSD:1315451.1315479} assumes the mapping between the utility and tuple's content is given, for example, by an application expert, while \cite{3:Tatbul:2003:LSD:1315451.1315479, 3:Olston:2003:AFC:872757.872825} learn this mapping online depending on the used query.    
The authors in \cite{3:Rivetti:2016:LSS:2933267.2933311} assume that all tuples have the same impact on QoR but tuples may have different processing latencies. Therefore, they drop those tuples that have the highest processing latencies. 
In \cite{Quoc:2017:SAC:3135974.3135989}, the authors fairly select tuples to drop from different input streams by combining two techniques, stratified sampling and reservoir sampling.
The authors in \cite{Tok:2008:SAP:1353343.1353414} also proposed  to use stratified sampling and reservoir sampling to perform approximate join.
In both these papers, the authors assume that tuples have the same utility values and impose the same processing latency which is, however, not true in CEP.  
In comparison to those approaches, we assume that there is a dependency between events in patterns and in input event streams in the context of CEP. Moreover, we assume that events might have different processing latencies.

In  \cite{3:He2014OnLS}, the authors proposed a load shedding  strategy for  CEP systems. They formulated the load shedding problem in CEP as a set of different optimization problems, where they consider a multi-pattern operator. The authors consider only the repetition of events in the input event stream and in patterns. However, they don't consider the order of events in both  the input event stream  and in patterns which is important in CEP, e.g., in sequence and negation operators. In \cite{eSPICE}, the authors proposed a load shedding strategy, called eSPICE, for CEP systems. eSPICE drops events from the operator's input event stream, where it considers the order and dependency of events in patterns and in the input event stream. It assigns utility values to events within a window where an event might have different utility values in different windows, depending on its position within the windows. eSPICE efficiently drops events that have the lowest utility values from windows by finding a utility value that can be used as a threshold utility to drop events. However, eSPICE imposes a higher overhead on the operator compared to pSPICE since it performs dropping on a finer granularity. Moreover, since eSPICE drops events, it might result in producing false positives, e.g., when using negation operator.

\section{Conclusion}
In this paper, we proposed an efficient, light-weight load shedding strategy, called \framework. In case of overload, \framework~drops PMs from  a CEP operator's internal state to maintain a given latency bound. 
To minimize the impact of load shedding on QoR, we proposed to utilize two important features (current state of a PM and number of remaining events in a window) that reflect the importance of PMs and used these features in calculating the utility of PMs, where we model the pattern matching operation as a Markov reward process.
By thoroughly evaluating \framework~with three real-world datasets and multiple important queries in CEP, we  show that \framework~considerably reduces  the degradation in QoR compared to state-of-the-art load shedding strategies. Moreover, we show that dropping partial matches instead of individual primitive events significantly reduces the overhead of load shedding on the system.  

\balance
\section*{Acknowledgement}
This work was supported by the German Research Foundation (DFG) under the research grant "PRECEPT II" (BH 154/1-2 and RO 1086/19-2).


%
%




\bibliographystyle{IEEEtran}
\bibliography{paper}

%

%
%

\end{document}